\documentclass[letterpaper,twocolumn,amsmath,amssymb]{revtex4}
\usepackage{graphicx}
\usepackage{dcolumn}
\usepackage{hhline}
\usepackage{psfrag}
\usepackage{color}
\usepackage{bm}

\begin{document}

\title{Crystalline Order On Riemannian Manifolds With\\Variable Gaussian Curvature And Boundary}

\author{Luca Giomi}
\email{lgiomi@physics.syr.edu}

\author{Mark Bowick}
\email{bowick@physics.syr.edu}

\affiliation{%
Department of Physics,
Syracuse University,
Syracuse, New York
13240-1130,
USA
}%

\begin{abstract}
We investigate the zero temperature structure of a crystalline
monolayer constrained to lie on a two-dimensional Riemannian
manifold with variable Gaussian curvature and boundary. A full
analytical treatment is presented for the case of a paraboloid of
revolution. Using the geometrical theory of topological defects in a
continuum elastic background we find that the presence of a variable
Gaussian curvature, combined with the additional constraint of a
boundary, gives rise to a rich variety of phenomena beyond that
known for spherical crystals. We also provide a numerical analysis
of a system of classical particles interacting via a Coulomb
potential on the surface of a paraboloid.
\end{abstract}

\maketitle

\section{\label{sec:intro}Introduction}

Crystalline structures are ubiquitous in nature because ordered
close-packed configurations frequently minimize the interaction
energy between the component units of the system. In soft condensed
matter systems, where curved or fluctuating geometries are
energetically accessible, the interplay between order and geometry
has been the subject of much attention in the last two decades
\cite{NelsonPeliti:1987,BowickNelsonTravesset:2000,BowickEtAl:2002,BauschEtAl:2003,DinsmoreEtAl:2002,Travesset:2005,ChushakTravesset:2005,BowickEtAl:2006,VitelliLucksNelson:2006}.
The existence of some preferred geometry, as for instance that
arising from the growth of a crystalline monolayer on a rigid
substrate, may influence the nature of the allowed order and, on the
other hand, the formation of particular ordered structure may lead
to a deformation in the geometry of the substrate (i.e. shape
transition
\cite{LidmarMirnyNelson:2003,NguyenBruinsmaGelbart:2005}). Self
assembled systems such as \emph{colloidosomes}
\cite{BauschEtAl:2003} or thin films of block copolymers
\cite{Kramer:2005} are realizations of such non-Euclidean ``soft''
crystals. Protein subunits, which comprise the shells of spherical
viruses, also provide an example in which the mechanical properties
of the systems are affected by the formation of crystalline
aggregates on surfaces equipped with a non-zero Gaussian curvature.

Some progress in understanding crystalline arrangements of particles
interacting on a curved surface was achieved by the authors of
\cite{BowickNelsonTravesset:2000,BowickTravesset:2001} in the
context of the geometric theory of topological defects. The original
interacting particle problem is mapped to a system of interacting
disclination defects in a continuum elastic curved background. The
defect-defect interaction is universal with the particle microscopic
potential determining two free parameters, the Young modulus of the
elastic background and the core energy of an elementary disclination
\cite{MembranesAndSurfaces}.

In flat two-dimensional space particles almost always pack in
triangular lattices, unless the interaction potential is carefully
tuned to select some other lattice topology. The addition of
non-zero Gaussian curvature to the problem introduces frustration in
the sense that perfect planar crystalline order is now incompatible
with the curvature of the surface. The most clear-cut example of
such \emph{geometrical frustration} is the case of spherical
crystals \cite{BowickEtAl:2002,BowickEtAl:2006}. The geometrical
frustration in terms of lattice topology is revealed by the Euler
theorem on the sphere which relates the number of vertices ($V$),
faces ($F$) and edges ($E$) of any convex polyhedron : $V-E+F=\chi$,
with $\chi=2$ in the case of the 2-sphere. This classical result of
topology can be rephrased in a form that is particularly useful to
describe the presence of defects in the lowest energy configuration
of a curved crystal by defining a \emph{topological charge} as the
departure from the ideal coordination number of a planar triangular
lattice: $q_{i}=6-c_{i}$, with $c_{i}$ the coordination number of
the $i-$th vertex. The Euler theorem states that the total
disclination charge of any spherical lattice must be equal to six
times the Euler characteristic $\chi$:
\begin{equation}\label{eq:spherical_charge}
Q=\sum_{i=1}^{V}q_{i}=6\chi=12.
\end{equation}
Although identical particles with hard core repulsion in a plane
typically pack into a triangular lattice with $6-$fold coordination,
any triangulation of the sphere is required to have twelve $5-$fold
disclinations (provided we restrict ourselves to the
energetically-preferred charge $q=\pm 1$ disclinations) as well as
an arbitrary number of $6-$fold vertices. These twelve extra
disclinations are the consequence of the geometrical frustration
associated the the curvature of the sphere.

Topological defects can appear in spherical lattices in the form of
isolated $5-$fold disclinations placed at the vertices of an
icosahedron, as frequently happens in the case of small viral
capsids, or grouped into one-dimensional arrays of dislocations
(tightly-bound $(5,7)-$fold disclination pairs). These arrays, also
known as grain boundary ``scars'', appear on the sphere when the
ratio $R/a$ ($R$ radius of the sphere, $a$ Euclidean lattice
spacing) exceeds a particular threshold value approximately equal to
5.

In this article we present an analytical and numerical study of the
defect structure of a two-dimensional crystal constrained to lie on
the surface of a paraboloid of revolution. The parabolic geometry
introduces two novel and important features compared to the sphere:
\emph{1}) a variable Gaussian curvature and \emph{2}) the presence
of a boundary. Both these features must be treated properly for a
thorough theoretical understanding.

The paper is organized as follows. In Sec. \ref{sec:geometry} we
briefly review the geometrical approach of
\cite{BowickNelsonTravesset:2000,BowickEtAl:2002,BowickEtAl:2006} in
which the basic degrees of freedom are the defects themselves rather
than the interacting particles and we derive the zero-temperature
energy of a paraboloidal crystal. This formalism has the advantage
of reducing the number of degrees of freedom as well as being rather
universal in the sense that it applies to a broad class of
interacting potentials. In Sec. \ref{sec:numerics} we discuss the
results obtained by the numerical minimization of the potential
energy of a system of classical charged particles interacting via a
Coulomb potential on the surface of a paraboloid in the light of the
geometrical approach. Sec. \ref{sec:conclusions} will be devoted to
conclusions.

\section{\label{sec:geometry}The geometrical approach}

\subsection{\label{sec:elastic_free_energy}The Elastic Free Energy}

Let $\mathbb{P}$ be the two-dimensional paraboloid of revolution in
$\mathbb{R}^{3}$ described in parametric form by:
\begin{equation}\label{eq:paraboloid}
\left\{
\begin{array}{l}
x = r\cos\phi\\
y = r\sin\phi\\
z = \frac{h}{R^{2}}\,r^{2}
\end{array}
\right.,
\end{equation}
where $h$ is the height of the paraboloid and $R$ the maximum
radius. In the following we will call $\kappa=2h/R^{2}$ the
normal curvature of the paraboloid at the origin. The metric
tensor $g_{ij}$ (with determinant $g$) and the Gaussian
curvature $K$ are given respectively by:
\begin{subequations}
\begin{equation}\label{eq:metric_tensor}
g_{ij} =
\left(
\begin{array}{cc}
1+\kappa^{2}r^{2} & 0\\
0 & r^{2}
\end{array}
\right),
\end{equation}
\begin{equation}\label{eq:gaussian_curvature}
K(r) = \frac{\kappa^{2}}{(1+\kappa^{2}r^{2})^{2}}\cdot
\end{equation}
\end{subequations}
Such paraboloidal surfaces serve as a good testing ground for
exploring the effects of both variable Gaussian curvature and the
presence of a boundary on the nature of crystalline order. From the
topological point of view a paraboloid of revolution is equivalent
to a disk. The Euler characteristic is thus $\chi=1$. The total
topological charge, taking into account the preferred coordination
on the boundary and in the interior, is given in this case by:
\begin{equation}\label{eq:parabolic_charge}
Q = \sum_{i=1}^{N_{b}}(4-c_{i})+\sum_{i=1}^{N_{i}}(6-c_{i}) = 6,
\end{equation}
where $N_{b}$ is the total number of disclinations on the boundary
and $N_{i}$ is the number of disclinations in the interior of the
paraboloid (in the following we will reserve the letter $N$ for the
total number of defects and $V$ for the total number of vertices of
the lattice). The elastic free energy of the crystal may be
expressed in the form
\cite{BowickNelsonTravesset:2000,BowickTravesset:2001,BowickEtAl:2002,BowickEtAl:2006}:
\begin{equation}\label{eq:free_energy}
F = F_{el} + F_{c} + F_{0},
\end{equation}
where $F_{0}$ is the free energy of the defect-free monolayer and
$F_{c}$ is the contribution to the free energy due to the core
energy of disclinations and is proportional to the total number of
defects. The elastic energy $F_{el}$ associated with the defect
interaction can be expressed in the form:
\begin{equation}\label{eq:elastic_free_energy}
F_{el} = \tfrac{1}{2}Y\int
d^{2}x\,d^{2}y\,G_{2L}(\bm{x},\bm{y})\rho(\bm{x})\rho(\bm{y}),
\end{equation}
where $Y$ is the Young modulus for the planar crystal. The quantity
$\rho(\bm{x})$ has the meaning of the effective topological charge
density:
\begin{equation}\label{eq:charge_density}
\rho(\bm{x}) = \frac{\pi}{3}\sum_{i=1}^{N}
q_{i}\delta(\bm{x},\bm{x}_{i}) - K(\bm{x}),
\end{equation}
where $\delta(\bm{x},\bm{x}_{i})$ is Dirac delta function on the
manifold:
$\delta(\bm{x},\bm{x}_{i})=g^{-1/2}\prod_{i}\delta(\bm{x}-\bm{x}_{i})$
and $G_{2L}(\bm{x},\bm{y})$ represents the Green's function for the
covariant biharmonic operator on $\mathbb{P}$.

The calculation of the effective free energy \eqref{eq:free_energy}
can be simplified if free boundary conditions are chosen. An application
of the second Green identity to Eq.\eqref{eq:elastic_free_energy} leads
straightforwardly to the form:
\begin{equation}\label{eq:free_boundary_energy}
F_{el} = \frac{1}{2Y}\int
d^{2}x\,\bigr[\Delta\chi(\bm{x})\bigr]^{2},
\end{equation}
where $\chi(\bm{x})$ is the solution of the inhomogeneous biharmonic equation:
\begin{equation}\label{eq:biharmonic_equation}
\Delta^{2}\chi(\bm{x}) = Y\rho(\bm{x})
\qquad\bm{x}\in\mathbb{P}
\end{equation}
with boundary conditions
\begin{subequations}
\begin{eqnarray}
\chi(\bm{x}) = 0                 &\quad \bm{x}\in\partial\mathbb{P}\label{eq:dirichlet} &\\
\nu_{i}\nabla^{i}\chi(\bm{x}) = 0 &\quad \bm{x}\in\partial\mathbb{P}\label{eq:neumann}   &,
\end{eqnarray}
\end{subequations}
in which $\nabla^{i}$ is the usual contravariant derivative in the
metric $g_{ij}$ and $\nu_{i}$ is the $i-$th component of the tangent
vector $\bm{\nu}$ perpendicular to boundary. If the parametrization
\eqref{eq:paraboloid} is chosen, the normal vector $\bm{\nu}$ is
simply given by $\bm{g}_{r}/|\bm{g}_{r}|$, with
$\bm{g}_{r}=\partial_{r}\bm{x}$ the base vector associated with
radial coordinate $r$. The solution of
Eq.\eqref{eq:biharmonic_equation} will then be:
\begin{equation}\label{eq:chi}
\chi(\bm{x}) = \int d^{2}y\,G_{L}(\bm{x},\bm{y})\Gamma(\bm{y}),
\end{equation}
where $G_{L}(\bm{x},\bm{y})$ is the Green's function of the
covariant Laplace operator on $\mathbb{P}$ with Dirichlet boundary
conditions
\begin{equation}\label{eq:green_equation}
\left\{
\begin{array}{ll}
\Delta G_{L}(\bm{x},\cdot) =\delta(\bm{x},\cdot)   &\quad\bm{x}\in\mathbb{P} \\[7pt]
G_{L}(\bm{x},\cdot)        = 0                     &\quad\bm{x}\in\partial\mathbb{P}
\end{array}
\right.
\end{equation}
and $\Gamma(\bm{x})=\Delta\chi(\bm{x})$ is the solution of the Poisson problem:
\begin{equation}\label{eq:poisson_equation}
\Delta\Gamma(\bm{x})=Y\rho(\bm{x}),
\end{equation}
which can be expressed in the Green form:
\begin{equation}\label{eq:gamma_1}
\frac{\Gamma(\bm{x})}{Y} = \int d^{2}y\,G_{L}(\bm{x},\bm{y})\rho(\bm{y})+U(\bm{x}),
\end{equation}
where $U(\bm{x})$ is an harmonic function on $\mathbb{P}$ that
enforces the Neumann boundary conditions \eqref{eq:neumann}.
\begin{figure}
\centering
{\psfrag{x}[c][c][1.2]{$r$}
 \psfrag{y}[r][r][1.2]{$\Gamma_{s}(r)$}
 \psfrag{1.}[c][c]{$1$}
 \psfrag{1.2}[c][c]{$1.2$}
 \psfrag{1.4}[c][c]{$1.4$}
 \psfrag{1.6}[c][c]{$1.6$}
 \psfrag{1.8}[c][c]{$1.8$}
 \psfrag{2.}[c][c]{$2$}
\includegraphics[scale=.75]{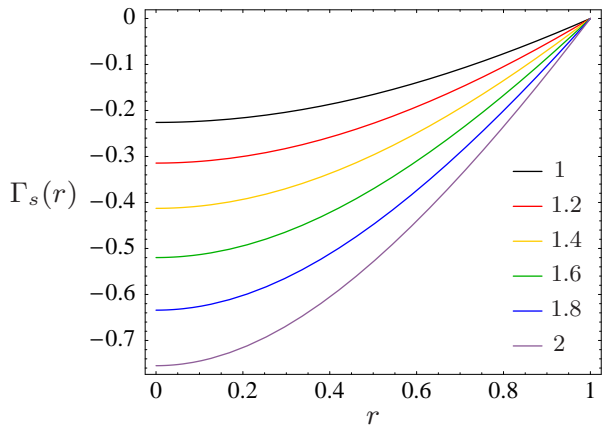}}
\caption{\label{fig:gamma_1}The function $\Gamma_{s}(r)$ for
different values of $\kappa$ in the range $1$-$2$.}
\end{figure}
The calculation of the Green's function $G_{L}(\bm{x},\bm{y})$
satisfying equation \eqref{eq:green_equation} can be reduced to the
more familiar planar problem by conformally mapping the paraboloid
$\mathbb{P}$ onto the unit disk of the complex plane (see Appendix
\ref{sec:green_function} for a detailed explanation):
\begin{equation}\label{eq:green_function}
G_{L}(\bm{x},\bm{y}) = \frac{1}{2\pi}\log\left|\frac{z(\bm{x})-z(\bm{y})}{1-z(\bm{x})\overline{z(\bm{y})}}\right|,
\end{equation}
where $z(\bm{x})=\varrho e^{i\phi}$, a point in the unit disk of the
complex plane, is the image of a point on the paraboloid under the
conformal mapping. The conformal distance $\varrho$ is related to
$r$ by:
\begin{equation}\label{eq:rho}
\varrho(r) = \lambda\frac{r e^{\sqrt{1+\kappa^{2}r^{2}}}}{1+\sqrt{1+\kappa^{2}r^{2}}}\,,
\end{equation}
with $\lambda$ a scale factor which ensures that $\varrho(R)=1$. As
explained in detail in Appendix \ref{sec:green_function}, the
Green's function $G_{L}(\bm{x},\bm{y})$, and hence the entire
elastic free energy, depends only on the coefficients of the first
fundamental form of the surface (i.e. the metric tensor $g_{ij}$).
Thus the elastic energy associated with the defect interactions and
thus the crystalline order is an \emph{intrinsic property} of the
manifold and so is invariant under local isometries. This
observation, which might appear obvious in the case of isometric
surfaces such the Euclidean plane and the cylinder, is quite
remarkable when applied to more sophisticated isometric manifolds
such as the catenoid and the helicoid or Scherk surfaces. We will
take this point up again in Sec. \ref{sec:conclusions}.

\begin{figure}[b]
\centering
{\psfrag{x}[c][c][1.2]{$r$}
 \psfrag{y}[r][r][1.2]{$U_{\kappa,R}$}
 \psfrag{1.}[c][c]{$1$}
 \psfrag{2.}[c][c]{$2$}
 \psfrag{3.}[c][c]{$3$}
 \psfrag{4.}[c][c]{$4$}
 \psfrag{5.}[c][c]{$5$}
 \psfrag{6.}[c][c]{$6$}
\includegraphics[scale=.75]{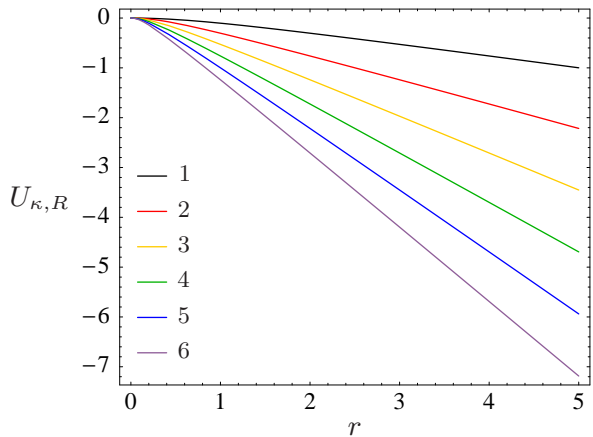}}
\caption{\label{fig:harmonic_constant}The quantity $U_{\kappa,R}$ as
a function of $R$ for different values of $\kappa$ in the range
$1$-$6$.}
\end{figure}

Inserting Eq.~\eqref{eq:green_function} into Eq.~\eqref{eq:gamma_1},
the function $\Gamma(\bm{x})$ can be expressed in the simple form:
\begin{equation}\label{eq:gamma_2}
\frac{\Gamma(\bm{x})}{Y} =
\frac{\pi}{3}\sum_{i=1}^{N}q_{i}G_{L}(\bm{x},\bm{x}_{i})-\Gamma_{s}(|\bm{x}|)+U(\bm{x}),
\end{equation}
where the first term represents the bare contribution of the defects
to the energy density and the second corresponds to the screening
effect of the Gaussian curvature. Explicitly:
\begin{equation}\label{eq:gamma_screening}
\Gamma_{s}(|\bm{x}|) = \log\left(\frac{\alpha e^{\sqrt{1+\kappa^{2}r^{2}}}}{1+\sqrt{1+\kappa^{2}r^{2}}}\right),
\end{equation}
where $r=|\bm{x}|$ and
\begin{equation}\label{eq:alpha}
\alpha = \frac{1+\sqrt{1+\kappa^{2}R^{2}}}{\exp\left(\sqrt{1+\kappa^{2}R^{2}}\right)}
\end{equation}
is a normalization constant depending on boundary radius $R$ and the
ratio $\kappa$.

Fig. \ref{fig:gamma_1} shows a plot of the screening function
$\Gamma_{s}(r)$ for different values of $\kappa$ in the range $1-2$.
As expected the contribution due to Gaussian curvature is
maximum at the origin of the paraboloid and drops to zero at the
boundary.

The calculation of the harmonic function $U(\bm{x})$
requires a little more effort. If the crystal was defect-free (or
populated by a perfectly isotropic distribution of defects) the
function $U(\bm{x})$ would be azimuthally symmetric and  constant on
the boundary. By the maximum principle of harmonic functions,
$U(\bm{x})$ would be then constant on the whole manifold and
dependent only on $\kappa$ and the radius $R$:
$U(\bm{x})=U_{\kappa,R}$. Such a constant can be determined by
integrating $\Delta\chi(\bm{x}) = \Gamma(\bm{x})$ and imposing the
boundary condition \eqref{eq:neumann}
\cite{VitelliLucksNelson:2006}. This gives:
\begin{equation}
U_{\kappa,R} = \frac{2\pi}{A}\int_{0}^{R} dr\,\sqrt{g}\,\Gamma_{s}(r),
\end{equation}
where $A$ is the area of the paraboloid:
\begin{equation}\label{eq:area}
A = \frac{2\pi}{3\kappa^{2}}\left[\left(1+\kappa^{2}R^{2}\right)^{\frac{3}{2}}-1\right].
\end{equation}
As shown in Figure \ref{fig:harmonic_constant}, the value of
$U_{\kappa,R}$ quickly approaches the linear regime as the size of
the radius increases:
\begin{equation}
U_{\kappa,R} \approx -\frac{1}{4}\,\kappa R +\frac{1}{3}\cdot
\end{equation}
Then for a defect-free configuration, the contribution of the
boundary to the energy density is a constant offset that persist
even for large radii. In the presence of disclinations, on the other
hand, the function $\chi(\bm{x})$ is no longer expected to be
azimuthally symmetric and the harmonic function $U(\bm{x})$ will not
be constant throughout the paraboloid. Introducing the harmonic
kernel $H(\bm{x},\bm{y})$ such that:
\begin{equation}\label{eq:harmonic_kernel}
U(\bm{x}) = -\int d^{2}y\,H(\bm{x},\bm{y})\rho(\bm{y}),
\end{equation}
the determination of $U(\bm{x})$ requires the calculation of the
function $H(\bm{x},\bm{y})$ associated with the Green's function of
the weighted biharmonic operator arising from the conformal map of
the paraboloid onto the unit disk of the complex plane. This problem
has been intensively studied in the past few years by the
mathematics community because it is connected with the extension of
the maximum principle for the weighted biharmonic operator of the
form $\Delta w^{-1} \Delta$ (see Hedenmalm \emph{et al.}
\cite{HedenmalmEtAl:1999} for a good survey on this topic). In the
case of radial weights, as arises from the conformal mapping of any
surface of revolution, the function $H(\bm{x},\bm{y})$ can be
calculated explicitly (see Shimorin \cite{Shimorin:1998}). For the
sake of completeness we report an exact expression of the harmonic
kernel $H(\bm{x},\bm{y})$ in Appendix \ref{sec:harmonic_kernel}. The
physical understanding of the solution \eqref{eq:gamma_2}, however,
doesn't require the complete solution. As shown from the numerical
data of Sec. \ref{sec:numerics}, the distribution of the defects
along the boundary is predominately symmetric, and thus we expect
the constant factor \eqref{eq:harmonic_kernel} to be the leading
contribution from the boundary even in the general case.
\begin{figure}[t]
\centering
\includegraphics[scale=.55]{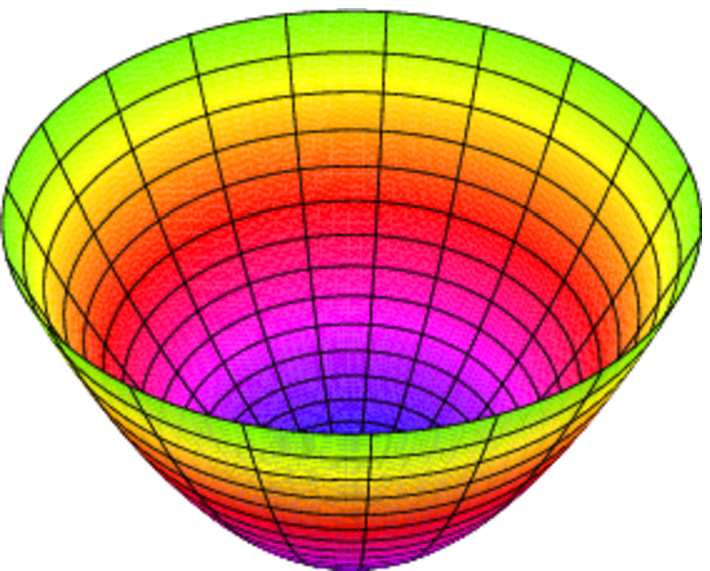}\qquad
\includegraphics[scale=.55]{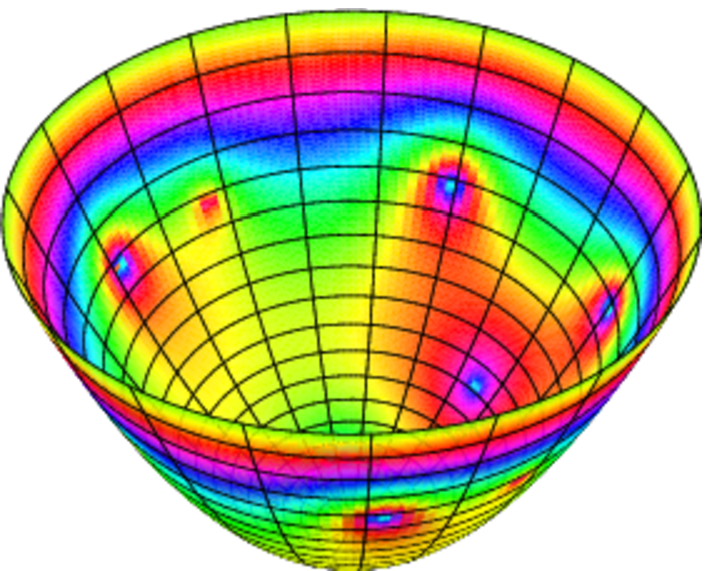}
\caption{\label{fig:colormap}Energy density $\Gamma^2(\bm{x})$ for
two different defect distributions. The energy density corresponding
to an isolated $+1-$disclination at the origin is shown on the left.
The defect distribution used for the figure on the right has been
obtained from the numerical minimization of the Coulomb energy of a
system of $V=50$ charges (see Sec. \ref{sec:numerics}).}
\end{figure}

For the ground state energy (zero temperature limit) that interests
us we must minimize the energy \eqref{eq:elastic_free_energy} with
respect to both the positions and the total number of defects. Since
the energy density \eqref{eq:gamma_2} depends on the difference
between a curvature term and the defect density we expect the
defects to arrange themselves so to approximately match the Gaussian
curvature. A complete screening of the Gaussian curvature would
yield a crystal with zero elastic energy at zero temperature. On the
other hand the core energy associated with a generic number $N$ of
disclinations will be linear in $N$ and so it will favor the fewest
defects possible. The structure of the crystal at zero temperature
will be governed therefore by the competition between the energy
cost of creating a defect and the compensating screening effect of
the Gaussian curvature. Once the distribution of the defects is
known the elastic energy can be easily calculated by integrating the
\eqref{eq:gamma_screening} on the whole paraboloid. In Fig.
\ref{fig:colormap} we show an example of the energy density
$\Gamma^{2}(\bm{x})$ corresponding to two different arrangements of
defects on a paraboloid with ($\kappa=1.6$ and $R=1$).

\subsection{\label{sec:large_core_energies}Large Core Energies: Pyramidal Lattices}

In the regime of large core energies $F_{c}\gg F_{el}$, the creation
of defects is strongly penalized and the lattice necessarily has the
minimum number of disclinations allowed by the topology of the
paraboloidal substrate. From symmetry considerations we might expect
the optimal distribution of defects to consist of $b$
$+1-$disclinations arranged along the boundary at the base vertices
of a $b$-gonal pyramid and a $b$-fold apex (of topological charge
$q_{0}=6-b$) at the origin. The homogeneous boundary conditions
adopted require the first term in Eq.~\eqref{eq:gamma_2} to vanish
when $\bm{x}_{i}\in\partial\mathbb{P}$. In the minimal energy
configuration then, the system has the freedom to tune the total
number of defects along the boundary to minimize the elastic energy
Eq.~\eqref{eq:free_boundary_energy} for any given value of the ratio
$\kappa$. This behavior is exclusive to manifolds with boundary and
doesn't have any counterpart in crystals on compact surfaces like
the sphere and the torus. In the following we will see how this
minimization leads to properties which we believe to hold, in the
most general sense, on any surface with boundary.
\begin{table}
\begin{ruledtabular}
\begin{tabular}{c|cccc}
$n$ & $Y_{3}$ & $Y_{4}$ & $Y_{5}$ & $Y_{6}$ \\
\hline
1  & 4   & 5   & 6   & 7   \\
2  & 10  & 13  & 16  & 19  \\
3  & 19  & 25  & 31  & 37  \\
4  & 31  & 41  & 51  & 61  \\
5  & 46  & 61  & 76  & 91  \\
6  & 64  & 85  & 106 & 127 \\
7  & 85  & 113 & 141 & 169 \\
8  & 109 & 145 & 181 & 217 \\
9  & 136 & 181 & 226 & 271 \\
10 & 166 & 221 & 276 & 331
\end{tabular}
\end{ruledtabular}
\caption{\label{tab:magic_number}Number of vertices $V$ for four
different $Y_{b,n}$ families in the range $b\in[3,6]$ (tetrahedron,
square, pentagonal and hexagonal pyramid) and $n\in[1,10]$.}
\end{table}

\begin{figure}[t]
\centering
\includegraphics[scale=0.3]{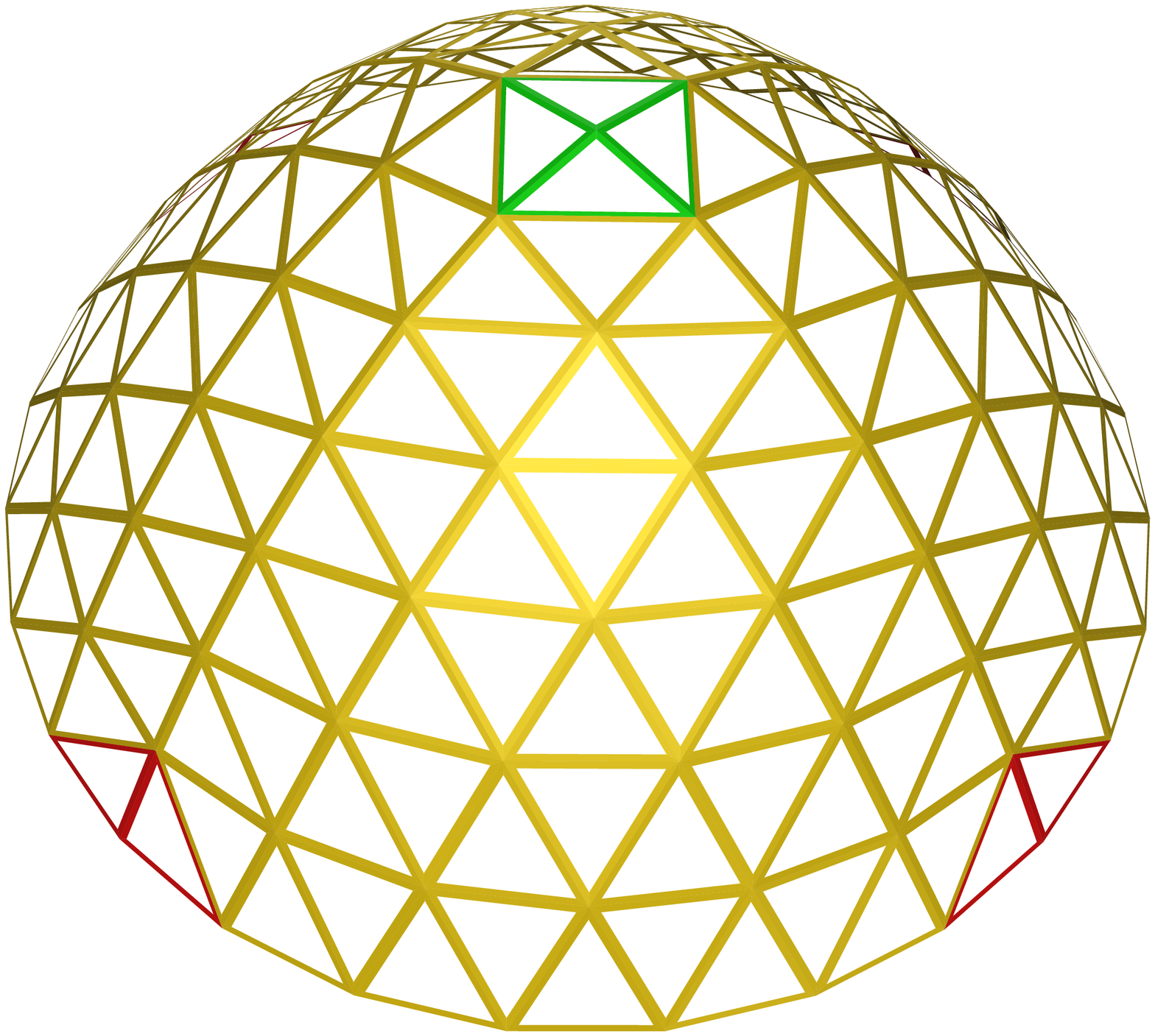}
\includegraphics[scale=0.3]{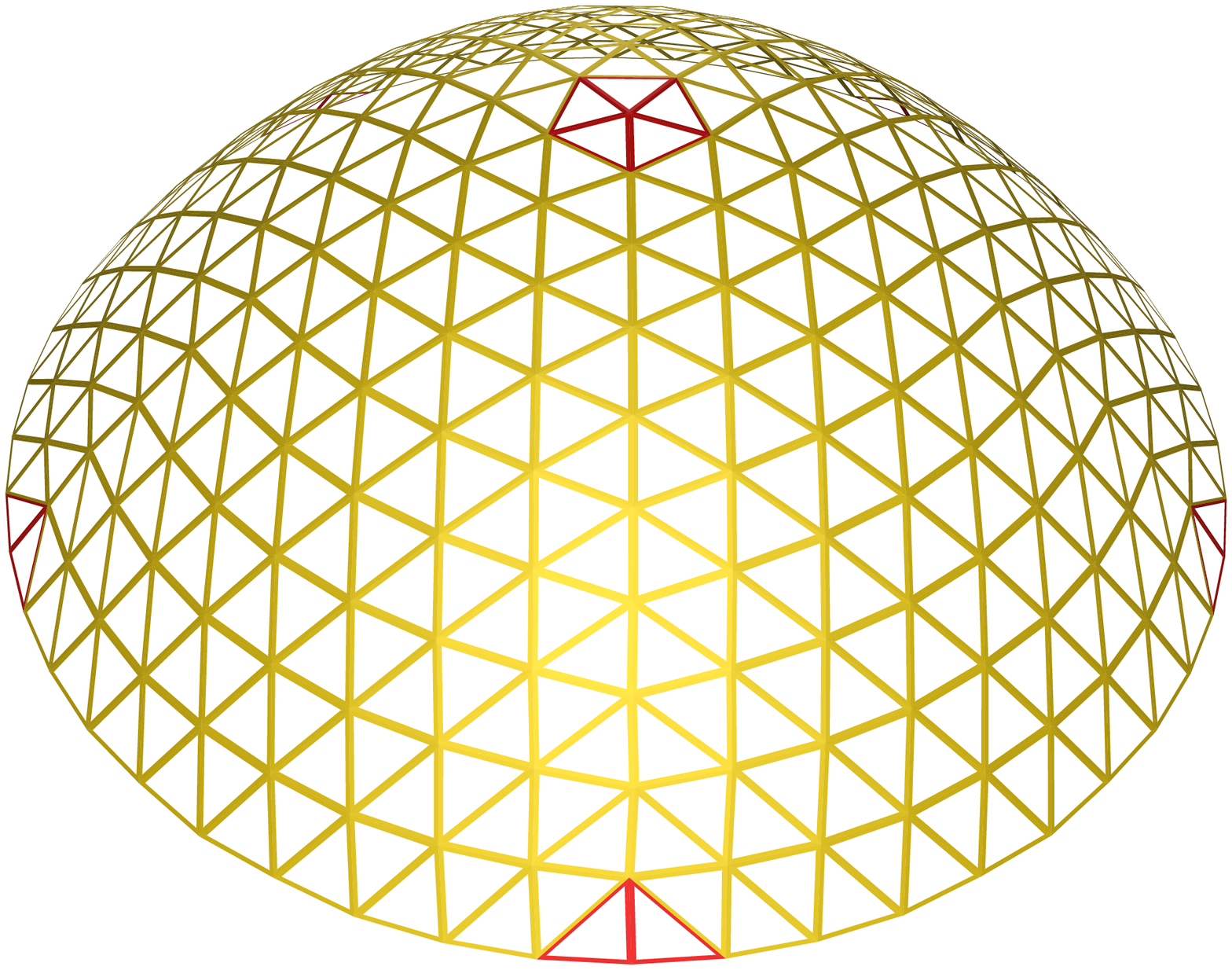}
\caption{\label{fig:pyramihedra}Two examples of $Y_{b,n}$
triangulations of the paraboloid ($Y_{4,7}$ on the top and
$Y_{5,10}$ on the bottom). Plaquettes with disclinations are
highlighted in red, for $+1-$disclinations, and green for
$+2-$disclinations.}
\end{figure}

\begin{figure}
\centering
{\psfrag{x}[c][c][1.2]{$R$}
 \psfrag{y}[r][r][1.2]{$\kappa$}
 \psfrag{y5}[c][c][1.2]{$Y_{5}$}
 \psfrag{y6}[c][c][1.2]{$Y_{6}$}
\includegraphics[scale=0.75]{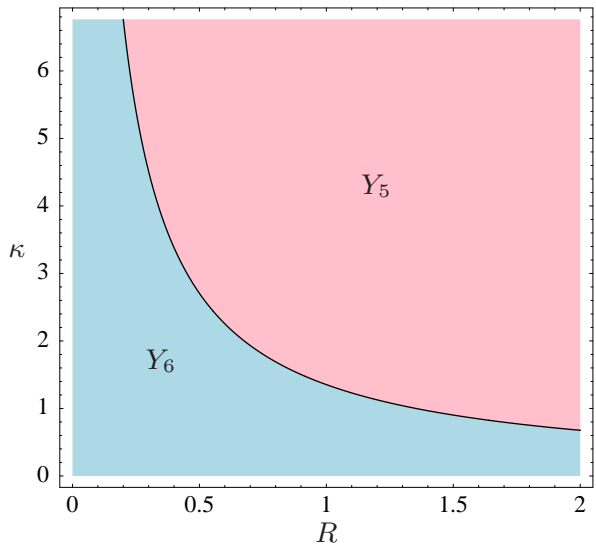}}
\caption{\label{fig:pyramid_phase_diagram}Phase diagram in the large
core energy regime. For small $\kappa$ the lattice preserves the
$6-$fold rotational symmetry of the flat case. As the curvature at
the origin increases the system undergoes a transition to the
$Y_{5}$ phase.}
\end{figure}

We will label a pyramidal configuration by $Y_{b}$, where $b$
denotes the number of base $+1-$disclinations. The coordinates
$(r,\phi)$ of the vertices are given by:
\begin{equation}\label{eq:pyramid_vertices}
Y_{b}: \qquad \left\{(0,\text{any}),\,\left(R,\frac{2\pi k}{b}\right)_{1\le k \le b}\right\}.
\end{equation}
Using the Euler theorem one can show that it is possible to
construct infinite families of polyhedra with the symmetry group
$C_{bv}$ from the pyramidal backbone $Y_{b}$. The number of vertices
is given by:
\begin{equation}\label{eq:pyramihedra}
V = \tfrac{1}{2}bn(n+1)+1,
\end{equation}
where $n$ is a positive integer which represents the number of edges
(not necessarily of the same length) of the polyhedron which
separates two neighboring disclinations. In the following we will
refer to these polyhedra with the symbol $Y_{b,n}$.
Fig. \ref{fig:pyramihedra} illustrates two $Y_{b,n}$ lattices for the
cases $b=4$ and $n=7$ (with $V=113$), and $b=5$ and $n=10$
($V=276$). In Table \ref{tab:magic_number} we report the number of
vertices for the four simplest $Y_{b,n}$ polyhedra for $n \in
[1,10]$.
By a numerical minimization of the energy
Eq.\eqref{eq:free_boundary_energy} one can establish that the
$Y_{b}$ are indeed equilibrium configurations for $b\in[3,5]$, for
some range of the parameters $\kappa$ and $R$. The cases of $b=5,6$
are particularly significant because they are characterized by an
equal number of defects ($N=6$) of the same topological charge
($q=1$). The two configurations will be associated therefore with
the same core energy $F_{c}$ and this introduces the possibility of a
structural transition between $Y_{5}$ and $Y_{6}$ governed by the
curvature ratio $\kappa$ and the boundary radius $R$. For fixed $R$
and small values of $\kappa$ the $6-$fold symmetric configuration
$Y_{6}$ is the global minimum of the free energy
Eq.\eqref{eq:elastic_free_energy}. For $\kappa$ larger than some
critical value $\kappa_{c}(R)$, however, the $Y_{6}$ crystal becomes
unstable with respect to the $5-$fold symmetric configuration
$Y_{5}$. A numerical calculation of the intersection point between
the elastic energies of $Y_{5}$ and $Y_{6}$ for different values of
$\kappa$ and $R$ allow us to construct the phase diagram shown in
Fig. \ref{fig:pyramid_phase_diagram}. The word ``phase'' in this
context refers to the symmetry of the ground state configuration as
a function of the geometrical system parameters $\kappa$ and $R$.

The scenario depicted in Fig. \ref{fig:pyramid_phase_diagram} can be
understood heuristically by imagining a system of spherically
symmetric equally sized subunits initially arranged on the surface
of a planar disk ($\kappa=0$). The most efficient packing of this
system is clearly the one in which the subunits are arranged in a
triangular lattice with six $3-$fold sites on the boundary at the
vertices of a hexagon. If now we slightly deform the disk into a
low-curvature paraboloid ($\kappa>0$) we might expect the hexagonal
configuration to persist for small values of $\kappa$. When the
deformation is more pronounced, however, the curvature at the origin
will be enough to support the existence of a $5-$fold vertex and the
system will undergo a structural transition from the $Y_{6}$ to the
$Y_{5}$ phase. In principle, if we keep increasing the curvature we
might expect the crystal to undergo a further transition to the
$Y_{4}$ phase.  In this case, however, the core energy will also
increase by a factor $4/3$ and so this is not generally possible in
the regime in which $F_{c} \gg F_{el}$. For intermediate regimes
(i.e. $F_{c} \sim F_{el}$), $Y_{5} \rightarrow Y_{4}$ and $Y_{4}
\rightarrow Y_{3}$ transitions are also possible. The critical value
of the parameters $\kappa$ and $R$, however, is not universal and
will depend in detail on the values of the core energy and the Young
modulus.

\subsection{\label{sec:small_core_energies}Small Core Energies: Scars And Coexistence}

When the core energy $F_{c}$ is small, the elastic energy
Eq.\eqref{eq:elastic_free_energy} can be lowered by creating
additional defects. In this section we present a simple argument for
the approximate phase diagram of a paraboloidal crystal in this regime.
A more detailed analysis of the minimal energy configurations
following from the solution Eq.\eqref{eq:gamma_2} will be reported
elsewhere.

Let us assume that a fivefold disclination is sitting at the point
$\bm{x}_{0}=(r_{0},\phi_{0})$ of $\mathbb{P}$. We can introduce a
notion of distance on the paraboloid by setting up a system of
geodesic polar coordinates $(s,\varphi)$ with origin at
$\bm{x}_{0}$. We expect that the stress introduced by the defect is
controlled by an effective disclination charge inside a circular
domain $C_{L}$ of geodesic radius $L$:
\begin{equation}\label{eq:effective_charge}
q_{eff}=q-\int_{0}^{2\pi} d\varphi\,\int_{0}^{L} ds\,\sqrt{g}\,K(s,\varphi),
\end{equation}
where $q=\pi/3$ is the charge of the isolated defect and the
integral measures the screening due to the total Gaussian curvature
within the domain. The metric tensor and the Gaussian curvature of a
generic Riemannian manifold can be expressed in geodesic polar
coordinates in the form (see for example Do Carmo \cite{DoCarmo}):
\begin{subequations}
\begin{gather}
g_{ij} =
\left(
\begin{array}{cc}
1 & 0\\
0 & G
\end{array}
\right),\\[7pt]
K(s,\varphi) = - \frac{\partial_{s}^{2}\sqrt{G}}{\sqrt{G}},
\end{gather}
\end{subequations}
where $G=\bm{g}_{\varphi}\cdot\bm{g}_{\varphi}$. Furthermore, an
expansion of the metric around the origin $(0,\varphi)$ yields:
\[
\sqrt{G} = s- \tfrac{1}{6} K_{0} s^{3} + o(s^{5}).
\]
For small distance from the origin, Eq.\eqref{eq:effective_charge}
becomes:
\begin{align}
q_{eff}
&= q+\int_{0}^{2\pi} d\varphi\,\int_{0}^{L} ds\,\partial_{s}^{2}\sqrt{G}\\[7pt]
&= q-\pi K_{0} L^{2} + o(L^{4}). \label{eq:charge_expansion}
\end{align}
The right hand side of Eq.\eqref{eq:charge_expansion} is a very
general expression for the effective disclination charge at small
distance and doesn't depend on the embedding manifold. If a grain
boundary is radiating form the original disclination, we expect the
spacing between consecutive dislocations to scale like $1/q_{eff}$,
with $a$ the lattice spacing \cite{BowickNelsonTravesset:2000}. When
$q_{eff} \rightarrow 0^{+}$ the dislocation spacing diverges and the
grain boundary terminates. Since the Gaussian curvature is not
constant on $\mathbb{P}$, the choice of the origin (i.e. the
position of the central disclination along the grain boundary)
affects the evaluation of $q_{eff}$. One can identify upper and
lower bounds by observing that:
\begin{subequations}
\begin{gather}
\max_{r} K(r) = K(0) = \kappa^{2},\\
\min_{r} K(r) = K(R) = \frac{\kappa^{2}}{(1+\kappa^{2}R^{2})^{2}}\cdot
\end{gather}
\end{subequations}
Unlike the case of surfaces with constant Gaussian curvature, we
expect the phase diagram for paraboloidal crystals to consist of
three regions separated by the curves:
\begin{equation}\label{eq:critical_length}
K_{0}L^{2} = \frac{1}{3}
\qquad
K_{0} = K_{\min},\,K_{\max}.
\end{equation}
When $L-L(K_{\min})\rightarrow 0^{+}$ the effective disclination
charge goes to zero and the distance between two consecutive
dislocations diverges at any point on $\mathbb{P}$. On the other
hand if $L-L(K_{\max})\rightarrow 0^{-}$, the disclination charge
will prefer to be delocalized in the form of grain boundary scars.
For $L(K_{\max})<L<L(K_{\min})$ the paraboloid will be equipped with
both regions where the Gaussian curvature is high enough to support
the existence of isolated disclinations and regions where the
screening due to the curvature is no longer sufficient and the
proliferation of grain boundary scars is energetically favored. This
leads to a three region phase diagram in which the regime of
isolated disclinations is separated from the delocalized regime of
scars by a novel phase in which both isolated disclinations and
scars coexist in different parts of the paraboloid according to the
magnitude of the Gaussian curvature.
\begin{figure}[t]
\centering
{\psfrag{x}[c][c][1.2]{$V$}
 \psfrag{y}[l][l][1.2]{$\kappa$}
 \psfrag{ID}[c][c][1.2]{ID}
 \psfrag{Coexistence}[c][c][1.2]{Coexistence}
 \psfrag{GrainBoundaries}[c][c][1.2]{Grain Boundaries}
\includegraphics[scale=.75]{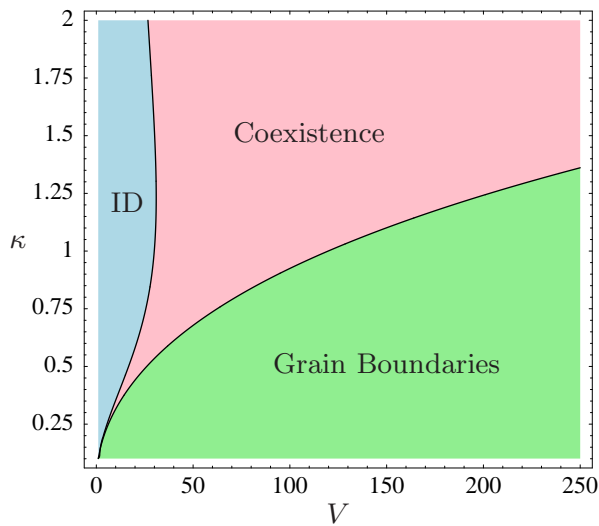}}
\caption{\label{fig:phase_diagram}Defect phase diagram for a
paraboloidal crystal of radius $R=1$. The two critical lines that
separate the isolated disclinations (ID) regime from the coexistence
regime and the coexistence regime from the scar phase correspond to
the solutions of Eq.\eqref{eq:critical_lattice_spacing} for
$K_{0}=K_{\min}$ and $K_{0}=K_{\max}$, respectively.}
\end{figure}

To compare this result with the structure of paraboloidal crystals
obtained by numerical minimization of the Coulomb energy (see Sec.
\ref{sec:numerics}) we need to measure the distance $L_{c}$ in terms
of the lattice spacing $a$ and rephrase
Eq.\eqref{eq:critical_length} as a condition on $a$ (or equivalently
on the number of vertices $V$). To do this we note that in order for
the domain $C_{L}$ to completely screen the topological charge of
the shortest scar possible (i.e. $5-7-5$), the geodesic radius $L$
has to be large enough to enclose the entire length of the scar.
Calling $\ell$ the geodesic distance associated with a single
lattice spacing $a$, we will then approximate $L\sim 3\ell$. In the
spirit of the local expansion Eq.\eqref{eq:charge_expansion}, we can
then approximate the neighborhood of the central point $\bm{x}_{0}$
with a spherical cap of radius $1/\sqrt{K_{0}}$. With this choice we
obtain:
\begin{equation}\label{eq:critical_lattice_spacing}
a^{2}\approx\frac{2}{K_{0}}\left[1-\cos\bigl(\ell\sqrt{K_{0}}\,\bigr)\right],
\end{equation}
which, combined with Eq.\eqref{eq:critical_length}, gives an
equation in $a$ and $\kappa$ for fixed $R$:
\begin{equation}\label{eq:critical_lines}
a^{2}\approx\frac{2}{K_{0}}\left(1-\cos\frac{1}{3\sqrt{3}}\right).
\end{equation}
The lattice spacing $a$ can be approximately expressed as a function
of the number of vertices of the crystal by dividing the area $A$ of
the paraboloid by the area of a hexagonal Voronoi cell of radius
$a/2$:
\begin{equation}\label{eq:lattice_spacing}
a^{2}\approx\frac{A}{\frac{\sqrt{3}}{2}\,V}\cdot
\end{equation}
The phase diagram arising from the solution of
Eq.\eqref{eq:critical_lattice_spacing}, together with the condition
Eq.\eqref{eq:lattice_spacing}, is sketched in Fig.
\ref{fig:phase_diagram} for the case $R=1$. The two critical lines
that separate the isolated defects (ID in the plot) regime from the
coexistence regime and this one form the grain boundaries phase
correspond respectively to the solutions for $K_{0}=K_{\min}$ and
$K_{0}=K_{\max}$. The simplicity of the criteria used to derive
Eqs.\eqref{eq:critical_lattice_spacing} and
\eqref{eq:lattice_spacing} doesn't allow us to predict the regions
surrounding the critical lines with high numerical accuracy, but
does provide a semi-quantitative picture of the novel phenomenology of
defects in non-Euclidean crystals that is generally supported by the
numerical results presented in Sec. \ref{sec:numerics}.

\section{\label{sec:numerics}Numerical Experiments}

\subsection{\label{sec:energy_minima}Energy Minima And Complexity}

In the following section we report the results of a numerical
minimization of a system of $V$ classical particles interacting via
a Coulomb potential $E=\sum_{i<j}1/|\bm{x}_{i}-\bm{x}_{j}|$ on the
surface of a paraboloid. The equilibrium configuration arising from
this optimization problem can be viewed as a direct realization of a
paraboloidal crystal and thus provides a testing ground for our
analytical results.

The determination of the equilibrium properties of complex systems
is complicated by the rich topography of the energy landscape, with
its many, often deep, local minima (valleys) separated by high
barriers (passes).  The number of local minima grows rapidly with
system size, making it increasingly difficult, or impossible, to
find the global minimum.

The effort in solving a given global optimization problem is
described by computational complexity theory. Locating the global
minimum for a potential energy surface belongs to the class of
problems known as NP-hard, for which there is no known algorithm
that is certain to solve the problem within a time that scales as a
power of the system size.

The Thomson problem
\cite{Thomson:1904,SaffKuijlaars:1997,HardinSaff:2005,ErberHockney:1995,MorrisDeavenHo:1996,AltschulerEtAl:1997,AltschulerGarrido:2005,WalesUlker:2006}
of finding the optimal configuration of $V$ interacting charges on a
2-sphere represents, in this context, a celebrated example of a hard
optimization problem. The existence of novel arrays of topological
defects in minimal energy configurations provides further insight
into the structure of the energy landscape. Computer experiments on
the Thomson problem indicate that, in the range $70 \le V \le 112$,
the number of local minima for each value of $V$ grows
exponentially: $\mathcal{N}\simeq 0.382\,\exp(0.0497\,V)$
\cite{ErberHockney:1995}. This trend is believed to continue for
larger values of $V$, making the determination of the global minimum
a formidable computational challenge. In the case of the paraboloid
we believe the prefactor in this scaling law is larger due to the
additional constraint of the boundary.

\begin{table}
\begin{ruledtabular}
\begin{tabular}{c|ccccc}
$V$ & $V_{-2}$ & $V_{-1}$ & $V_{0}$ & $V_{1}$ & Energy\\
\hline\\[-7pt]
10  & 0  & 0 & 4 & 6  & 36.94485696974016\\[2pt]
20  & 0  & 4 & 6 & 10 & 179.5291483377297\\[2pt]
30  & 0  & 6 & 12 & 12 & 439.0497473530407\\[2pt]
40  & 2  & 5 & 18 & 15 & 818.8300625504069\\[2pt]
50  & 4  & 4 & 24 & 18 & 1321.878894548272\\[2pt]
60  & 2  & 10 & 28 & 20 & 1949.230291403783\\[2pt]
70  & 2 &  16 &  26 &  26 &  2701.959660541221\\[2pt]
80  & 3 &  17 &  31 &  29 &  3581.110585181344\\[2pt]
90 &  2 &  16 & 46 &  26  & 4588.364706108566\\[2pt]
100 &  3 &  15 &  55 &  27 &  5722.503370970009\\[2pt]
150 &  1 &  30 &  81 &  38 &  13323.70617345018\\[2pt]
200 &  3 &  35 &  115 &  47 &   24173.21580330549l
\end{tabular}
\end{ruledtabular}
\label{tab:numerical_data} \caption{Numerical data for twelve selected
lattices. The quantities $V_{q}$ represent the number of vertices in
the crystal with topological charge $q$.}
\end{table}

\begin{figure}
\centering
\includegraphics[height=120pt]{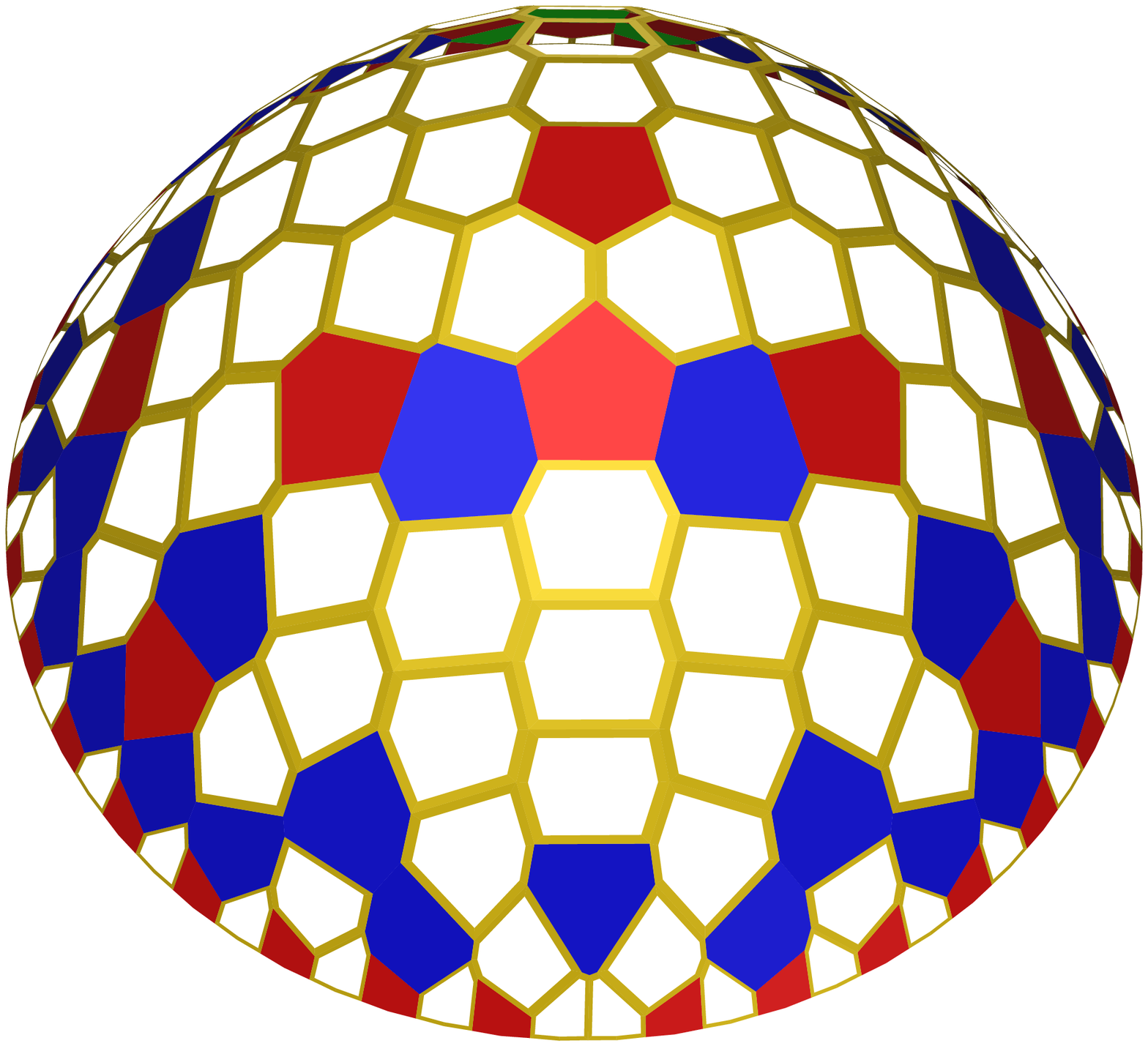}
\includegraphics[height=120pt]{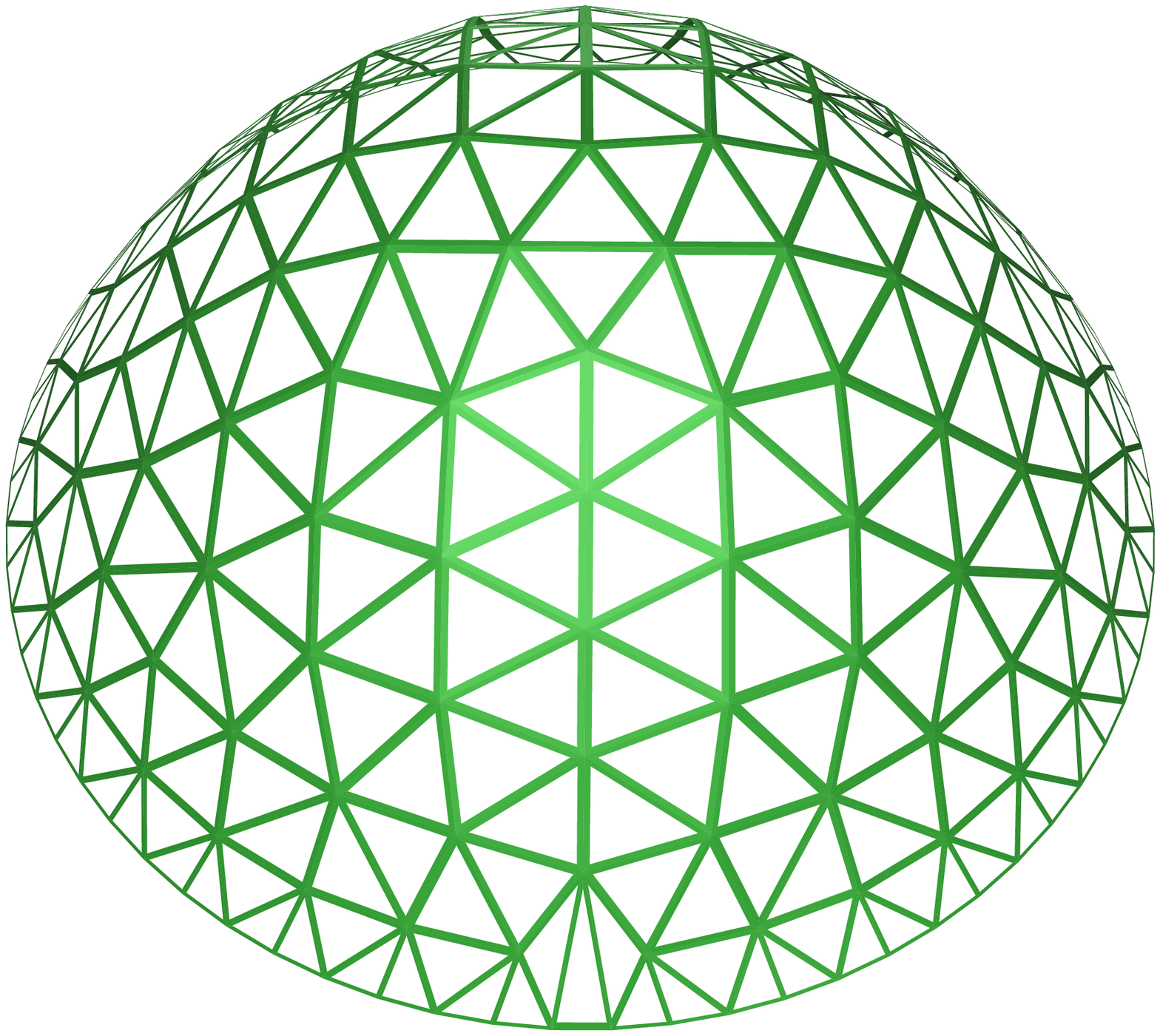}\\
\includegraphics[height=120pt]{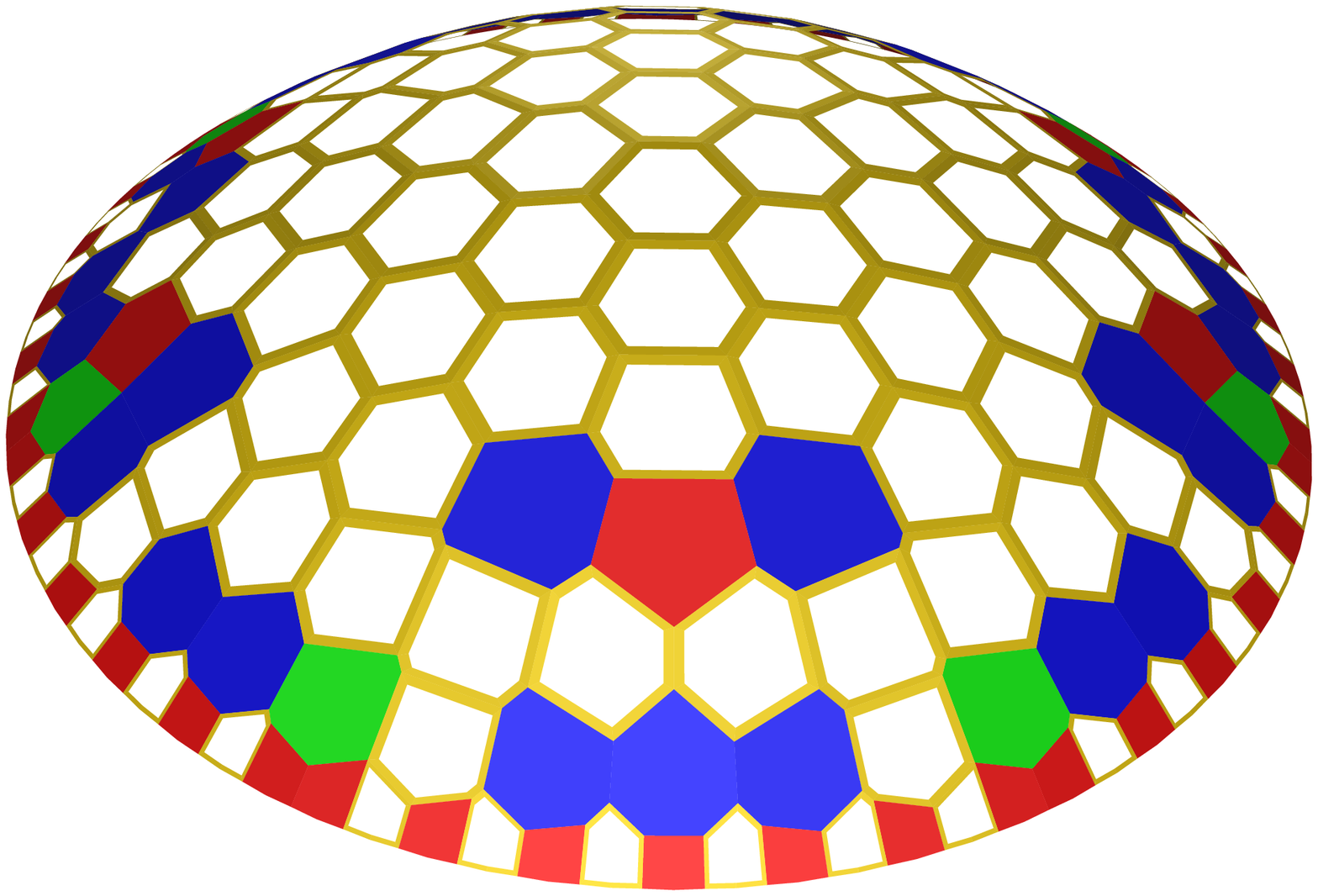}
\includegraphics[height=120pt]{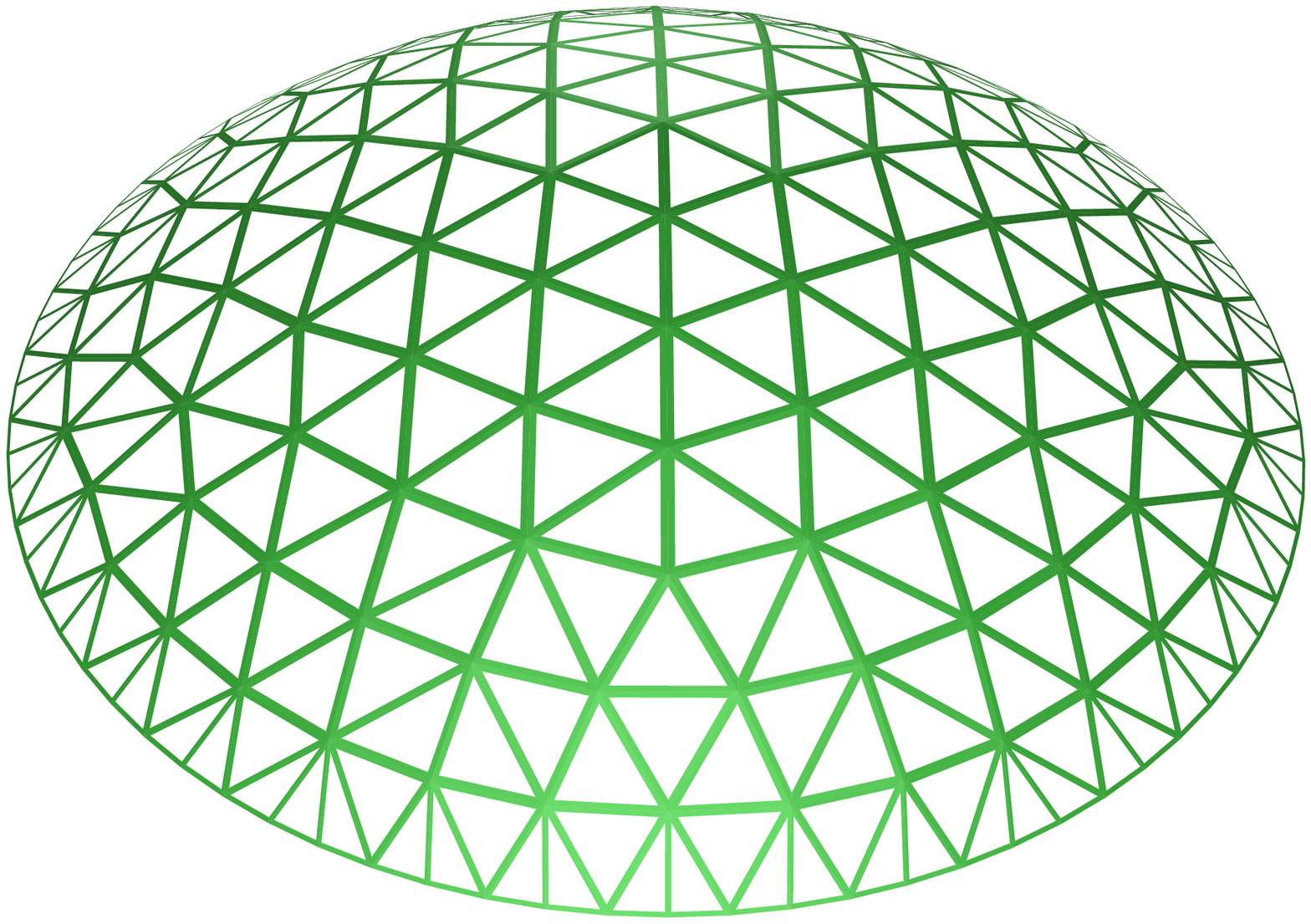}\\
\includegraphics[height=80pt]{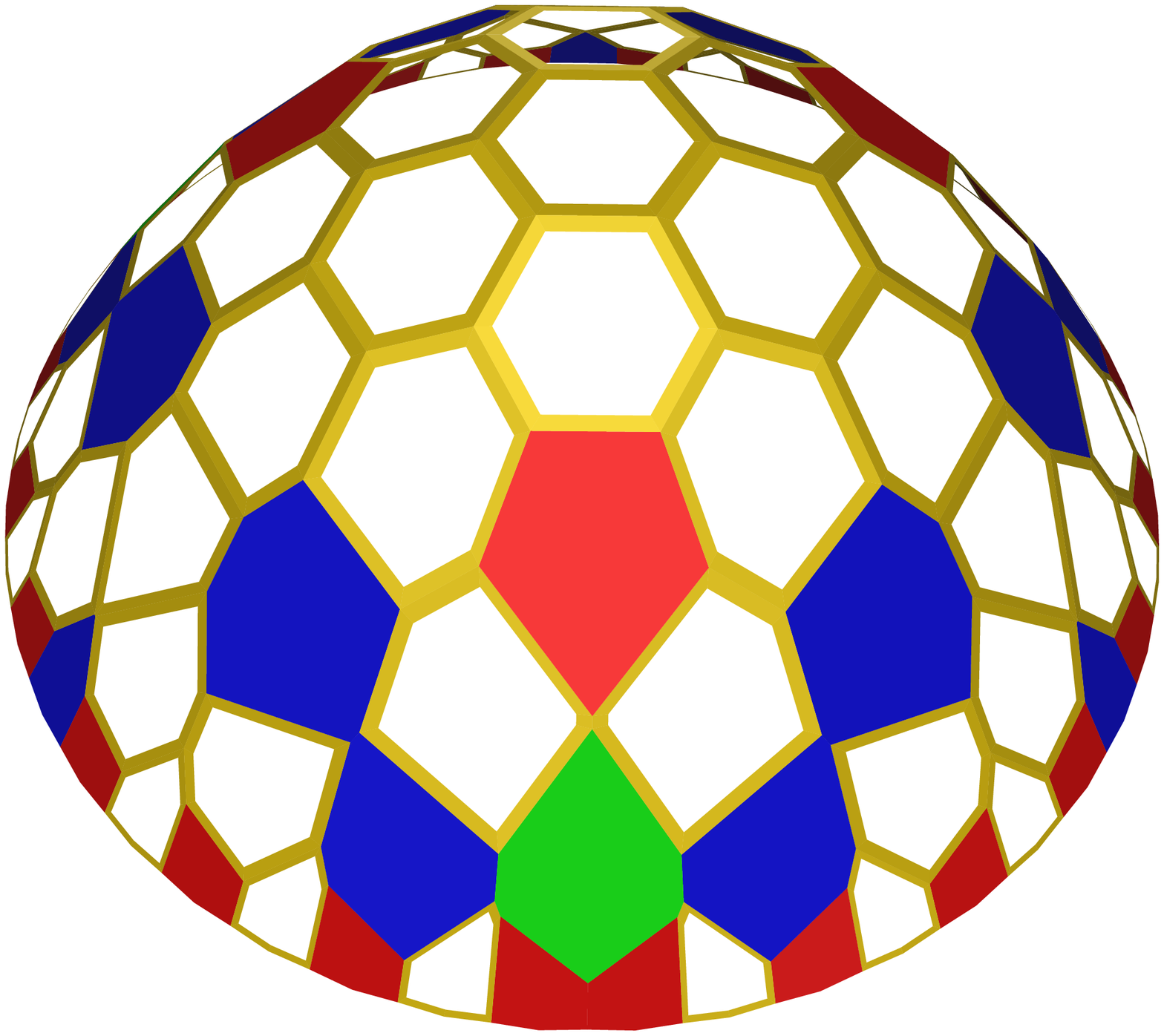}
\includegraphics[height=80pt]{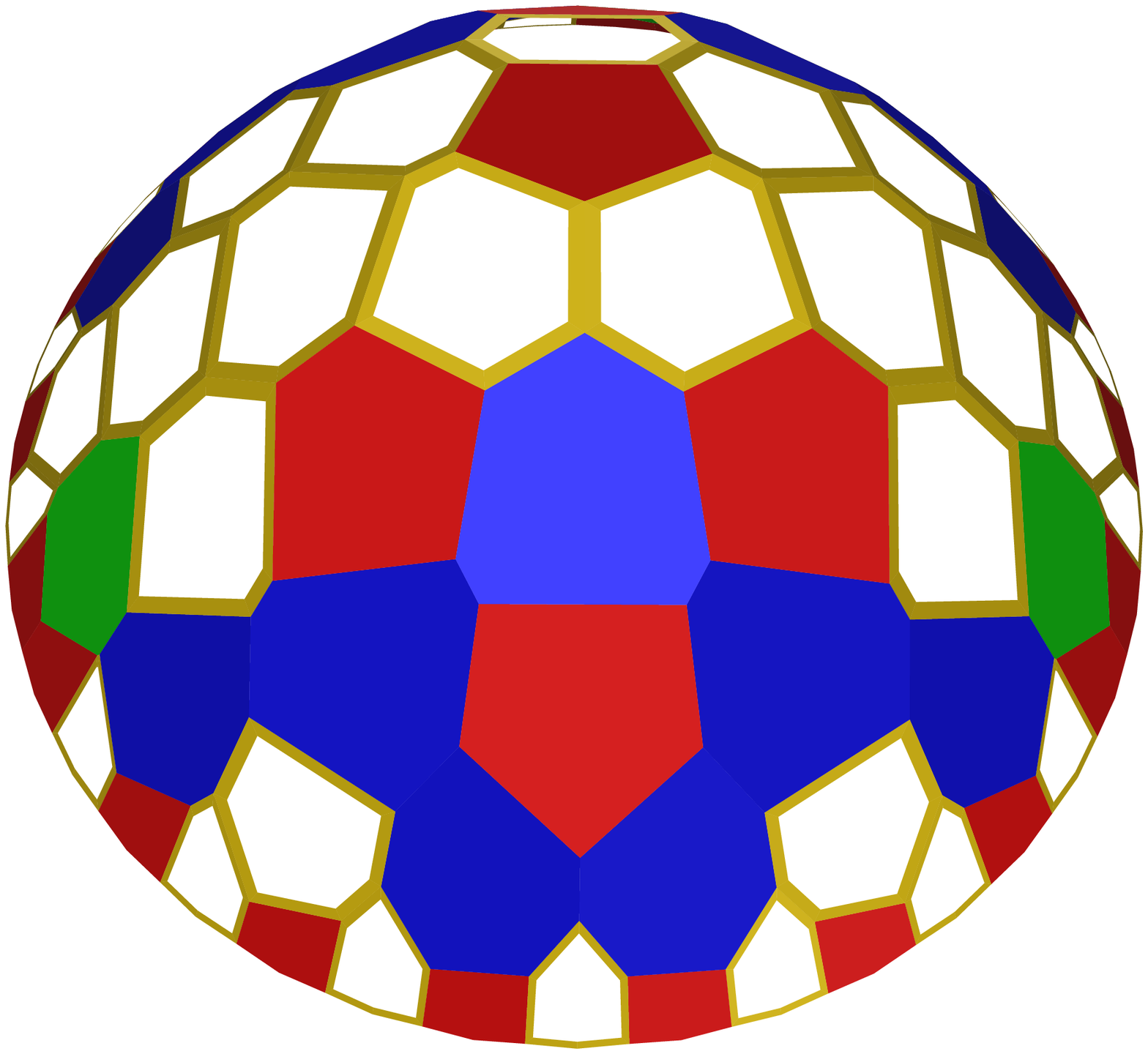}
\includegraphics[height=80pt]{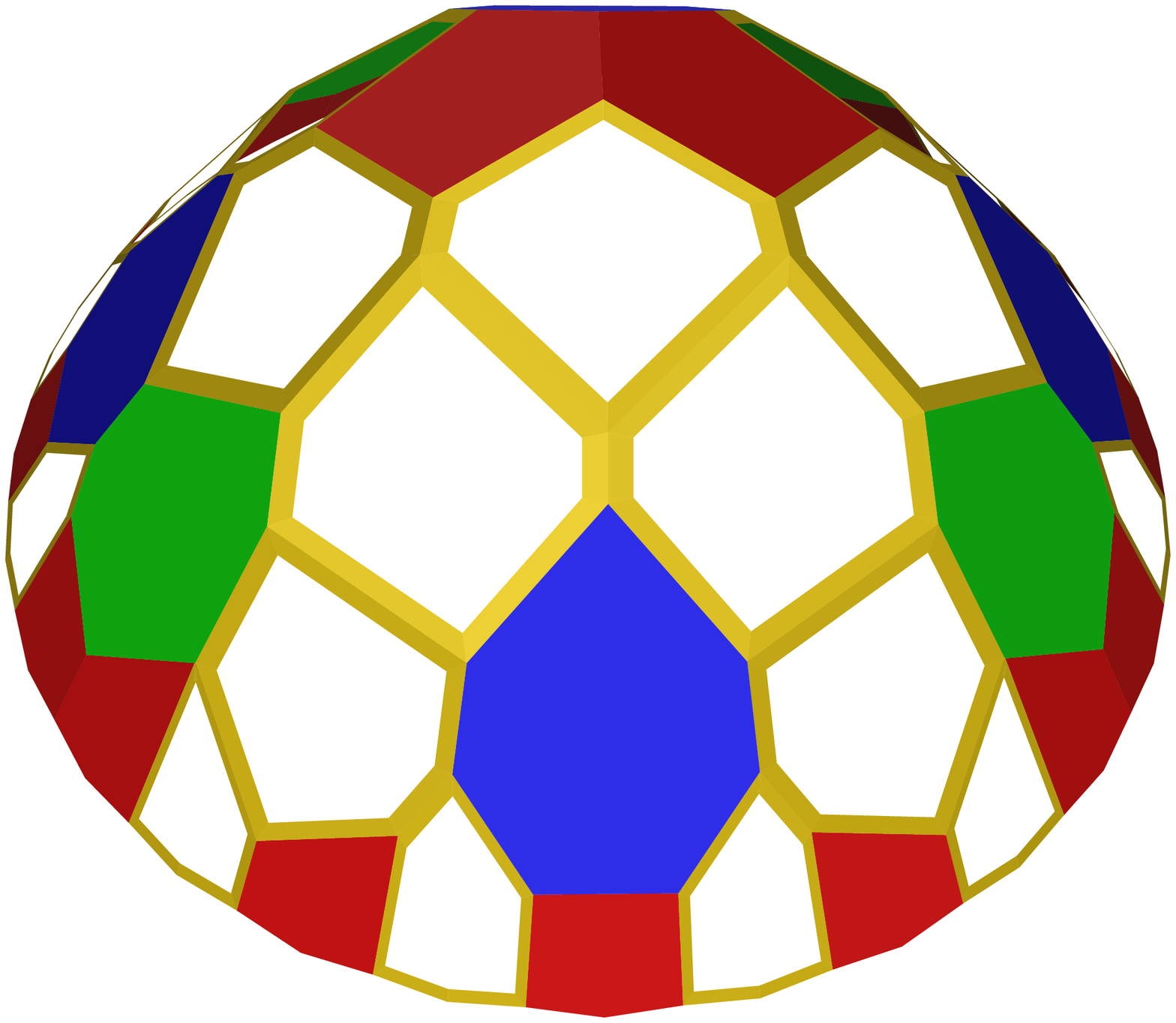}\\
\includegraphics[height=80pt]{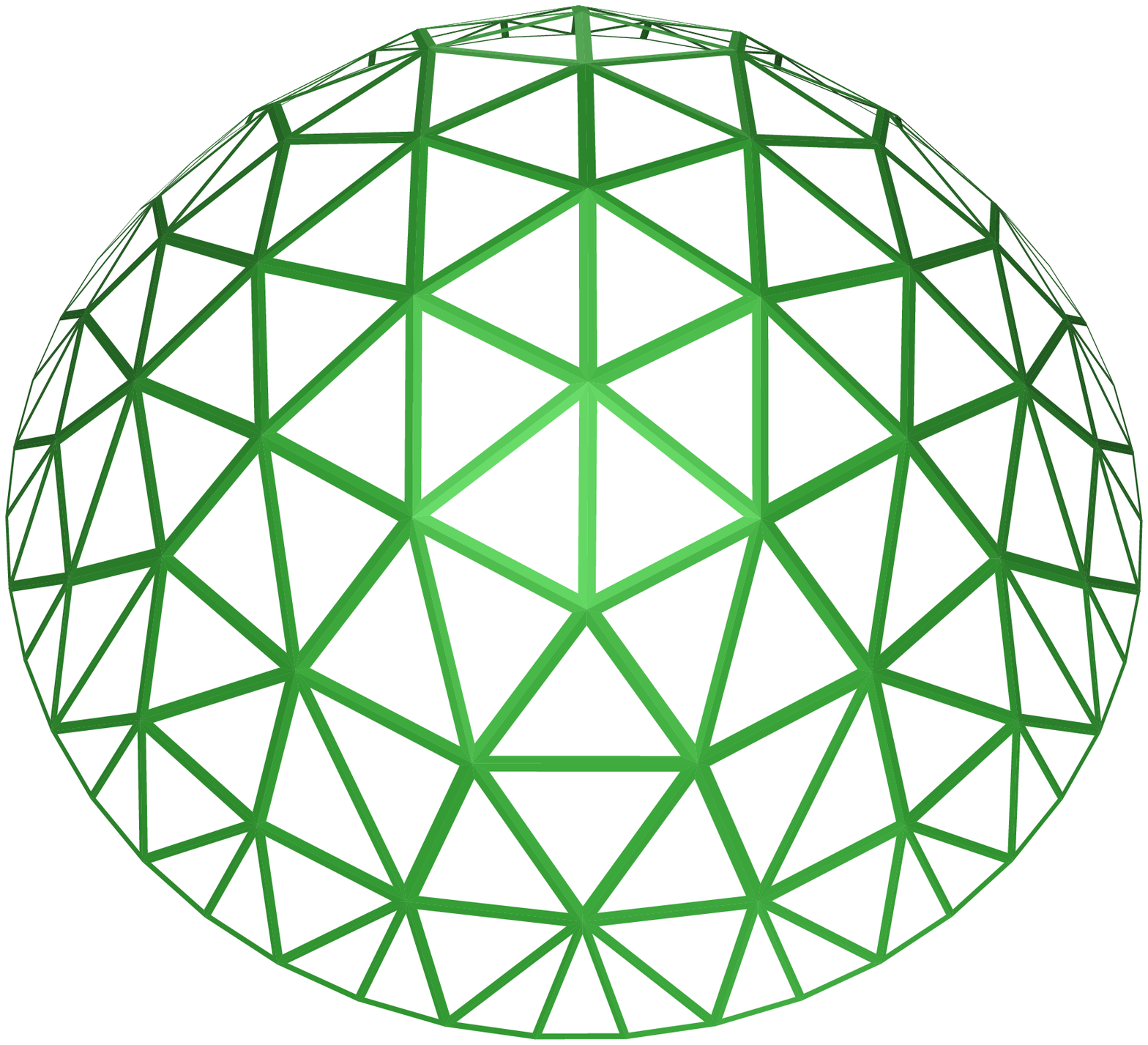}
\includegraphics[height=80pt]{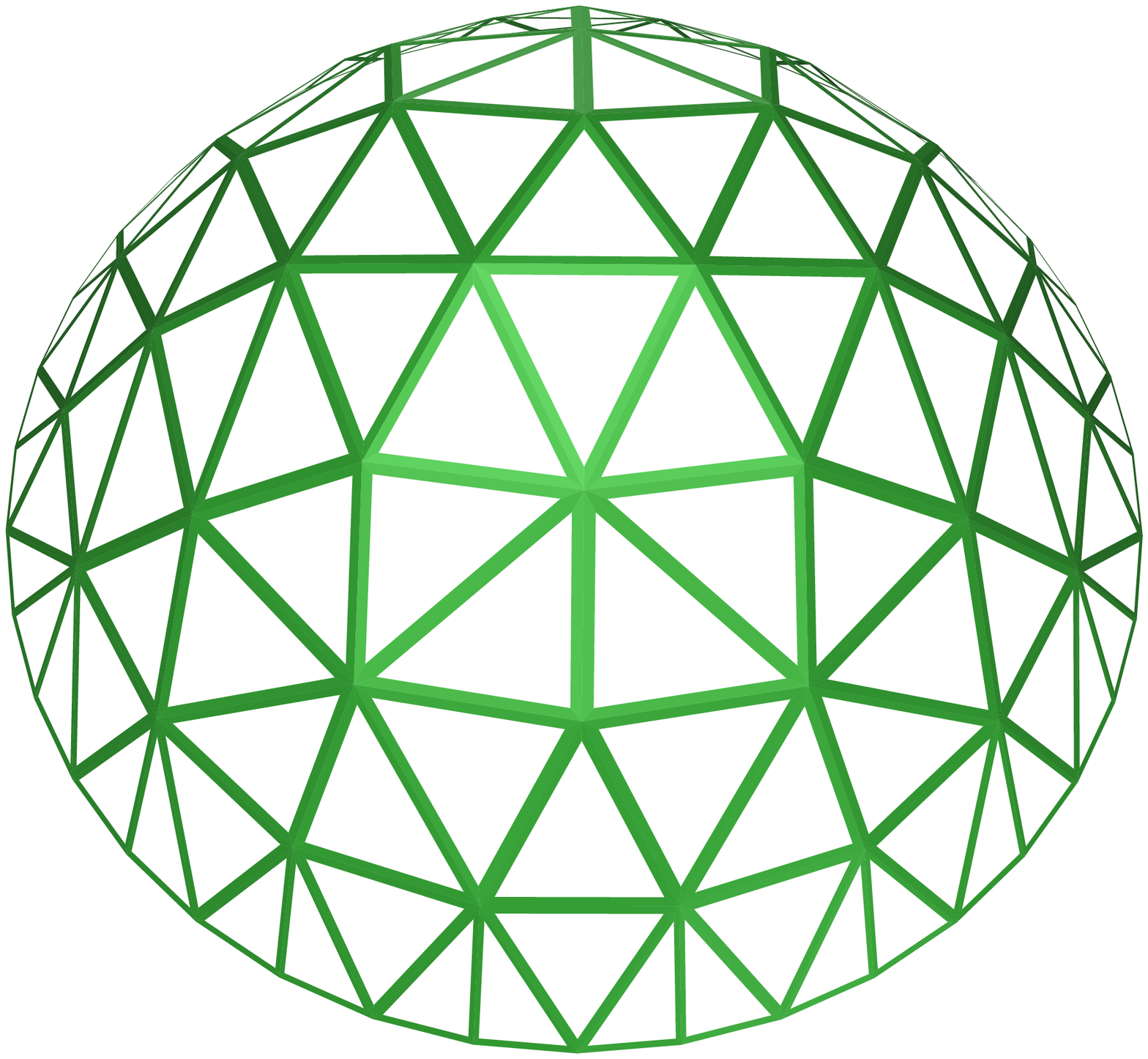}
\includegraphics[height=80pt]{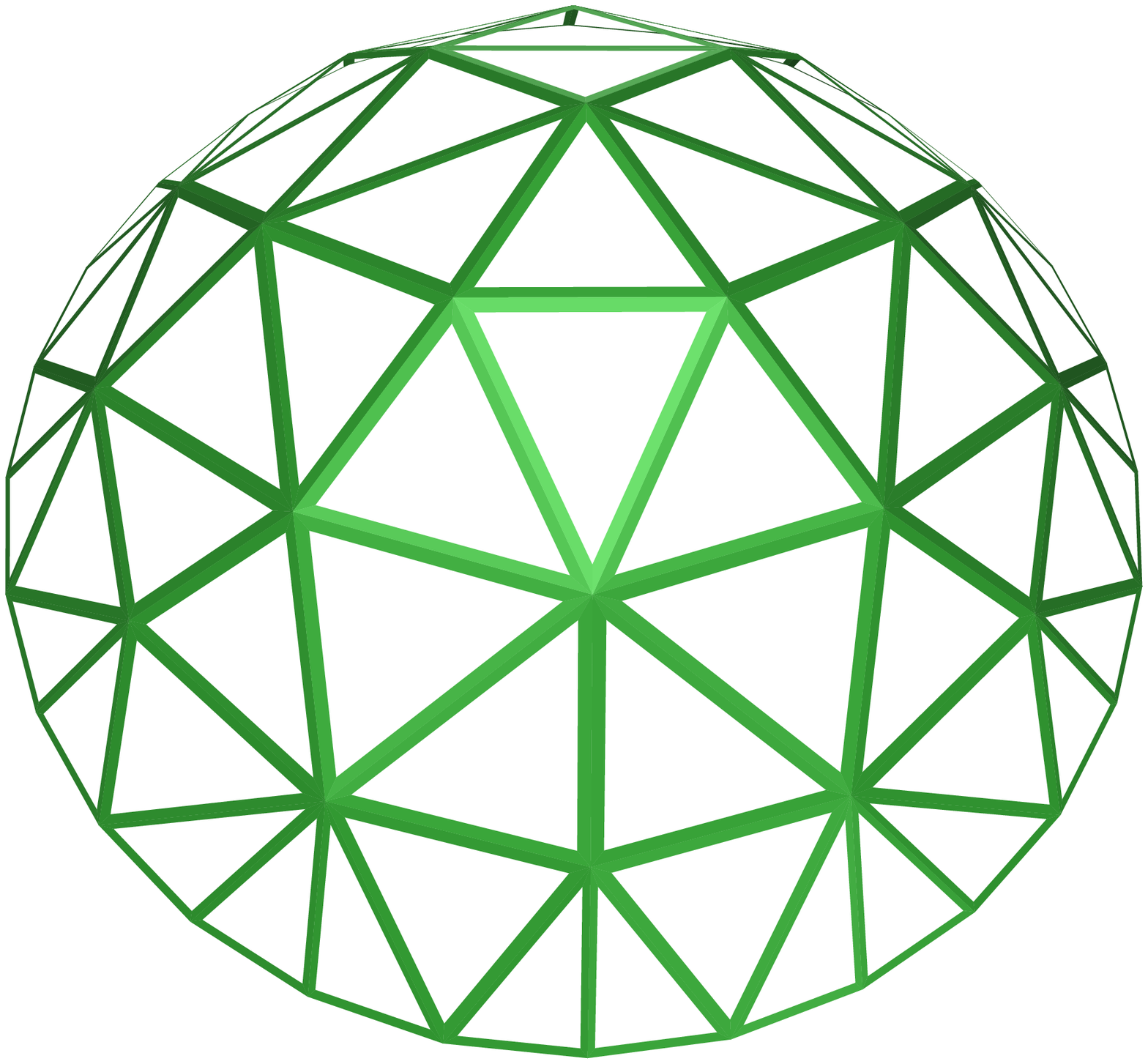}
\caption{\label{fig:parabolic_crystals}Voronoi lattices and Delaunay
triangulations for five selected systems from numerical simulations
with $R=1$. The first row corresponds to $V=200$ and $\kappa=1.6$,
while the second row is for $V=200$ and $\kappa=0.8$ (see Sec.
\ref{sec:parabolic_crystals} for a discussion). In the bottom two
rows $V=100,\,80,\,50$ with $\kappa=1.6$.}
\end{figure}

\subsection{\label{sec:parabolic_crystals}Paraboloidal Coulomb Crystals}

To construct equilibrium lattice configurations we adopted a
parallel implementation of Differential Evolution (DE) \cite{Code}
(see App. \ref{sec:differential_evolution} for an introductory
description of the algorithm together with the parallelization
strategy). The initial pool of candidate solutions is generated at
the beginning of the simulation by randomly creating $NP=20V$
configurations uniformly distributed over the whole search space
$\{r\in[0,R]\}\otimes\{\phi\in[0,2\pi]\}$. The population is then
evolved by $3 \cdot 10^{5}$ DE iterations on ten processors working
in parallel.

In Fig. \ref{fig:parabolic_crystals} we show the Voronoi lattice and
the Delaunay triangulations for five selected systems up to $V=200$
particles. The lowest energy seen, together with the number of
$n-$fold vertices, for each one of these lattices is reported in
Table \ref{tab:numerical_data}. In all the systems observed
disclinations always appear clustered in either grain boundary scars
or dislocations with the exception of isolated $+1-$disclinations
which appearing in the bulk as expected from the curvature screening
argument discussed in Sec. \ref{sec:geometry}. The complex
aggregation of defects along the boundary together with the presence
of negatively charged clusters indicates that the effect of the
boundary, in the case of relatively small systems like the ones
simulated, is more drastic than predicted by the homogeneous
boundary conditions Eq.\eqref{eq:dirichlet} and
Eq.\eqref{eq:neumann}. Even in the computationally expensive case of
$V=100$, the distance between the origin and the boundary of the
paraboloid is only four lattice spacings. In this situation we
expect the distribution of particles along the boundary to play a
major role in driving the order in the bulk.

For larger systems, such as $V=200$ (top of Fig.
\ref{fig:parabolic_crystals}), the behavior of the particles in the
bulk is less affected by the boundary and the crystalline order
reflects more closely the free-boundary problem discussed in Sec.
\ref{sec:geometry}. A comparison of the lattices in the first two
rows of Fig. \ref{fig:parabolic_crystals}, in particular, reveals
substantial agreement with the scenario described in
Sec. \ref{sec:large_core_energies}. For $\kappa=0.8$ and $V=200$, the
defects are all localized along the boundary with the exception of
one length-$3$ scar in the bulk at distance $r \approx 0.63$ from
the center. For $\kappa=1.6$, the pattern of defects in the bulk is
characterized by the coexistence of an isolated $+1-$disclination at
the origin and a length-$5$ $W-$shaped scar displaced along a
parallel one lattice spacing away from the central disclination.
Apart from the evident difficulty in comparing the structures of
small systems with those predicted from continuum elasticity theory,
this behavior is consistent with the simple picture sketched in the
phase-diagram of Fig. \ref{fig:phase_diagram}. The local $5-$fold
symmetry at the origin of the $\kappa=1.6$ configuration, compared
with $6-$fold symmetry for $\kappa=0.8$, suggests, as in the case of
spherical crystals \cite{EinertEtAl:2005}, that the complicated
structure of defect clusters appearing in large systems is the
result of the instability of the simpler $Y_{b,n}$ configurations
from which they partially inherit their overall symmetry. A more
accurate numerical verification of our theory remains a challenge
for the future.

The symmetry of the configurations presented in Fig.
\ref{fig:parabolic_crystals} deserves special attention. As for any
surface of revolution, the circular paraboloid belongs possesses the
symmetry group $O(2)$ of all rotations about a fixed point and
reflections in any axis through that fixed point. Any given
triangulation of the paraboloid may destroy the full rotational
symmetry completely or just partially, leaving the system in one of
the following two subgroups: the pyramidal group $C_{nv}$ or the
reflection symmetry group $C_{s}$. In general we found the latter
symmetry group for system sizes up to $V=200$ particles. The
symmetry for larger system sizes is under investigation.

\section{\label{sec:conclusions}Conclusions And Discussions}

We have analyzed both analytically and numerically the structure of
a two-dimensional paraboloidal crystal as a specific realization of
the class of crystalline structures on two-dimensional Riemannian
manifolds with variable Gaussian curvature and boundary. Using the
geometrical approach developed in \cite{BowickNelsonTravesset:2000}
we found that the presence of variable Gaussian curvature, combined
with the presence of a boundary, gives rise to a rich variety of
phenomena which we believe to be generic. The freedom of tuning the
optimal number of disclinations to place on the boundary, where the
presence of defects doesn't cost additional energy, allows the
crystal to undergo a structural transition controlled by the value
of the maximum Gaussian curvature of the surface, in the regime
where the creation of defects is energetically disfavored and
disclinations are isolated in the crystal and far apart from each
other. When the core energy of the defects is much smaller than the
elastic energy and disclinations are allowed to proliferate, the
presence of a variable Gaussian curvature is responsible for the
existence of an intermediate phase in which both isolated defects
and grain boundary scars coexist in the crystal in different regions
according to the magnitude of the curvature.

By conformally mapping the paraboloid into the unit disk of the
complex plane, we showed furthermore that the elastic energy of the
crystal only depends on the coefficients of the first fundamental
form of the surface and thus is an intrinsic property of the
manifold which is invariant for local isometries. This property,
which is somehow already contained in the linearity of the elastic
theory adopted, discloses a deep and fascinating feature of curved
crystals by requiring isometric surfaces (as the catenoid and the
helicoid or the Scherk surfaces) to support the same
crystalline order. From the other hand this observation can be used
to set a bound in the completeness of our theory and the accuracy of
linear elasticity in describing the physics of defects in curved
crystals. We hope more work will be devoted to the understanding of
these phenomena in the future.

\acknowledgments
This work was supported by the NFS through Grant No. DMR-0219292
(ITR) and through founds provided by Syracuse University.

\appendix

\section{\label{sec:green_function}Evaluation of the functions $G_{L}(\bm{x})$ and $\Gamma(\bm{x})$}

In this appendix we discuss the calculation of $\Gamma(\bm{x})$ in
Eq.\eqref{eq:gamma_2}. Consider the solution of the Green equation
\begin{equation}\label{eq:biharmonic_green_equation}
\Delta^{2}G_{2L}(\bm{x},\bm{y}) = \delta(\bm{x},\bm{y})
\qquad \bm{x},\,\bm{y} \in\mathbb{P}
\end{equation}
with homogeneous boundary conditions. In integral form this solution
can be written:
\begin{equation}\label{eq:biharmomic_green_function}
G_{2L}(\bm{x},\bm{y}) = \int d^{2}z\,G_{L}(\bm{x},\bm{z})[G_{L}(\bm{z},\bm{y})-H(\bm{z},\bm{y})]
\end{equation}
where $G_{L}(\bm{x},\bm{y})$ is the solution of the Green-Dirichlet
problem Eq.\eqref{eq:green_equation} and $H(\bm{x},\bm{y})$ is the
reproducing kernel of Eq.\eqref{eq:harmonic_kernel}. As noted in
Sec. \ref{sec:elastic_free_energy}, one can map the paraboloid
$\mathbb{P}$ onto the unit disk of the complex plane and then employ
the appropriate planar techniques (i.e. image charges). In general
any simply connected two-dimensional Riemannian manifold with a
$C^{\infty}-$smooth metric $ds^{2}$ can be equipped with a set of
local \emph{isothermal} (or conformal) coordinates $(x,y)$ such that
the metric is represented in the form $ds^{2}=w(x,y)(dx^{2}+dy^{2})$
for some positive conformal weight $w$. In two dimensions this local
system of isothermal coordinates can serve as a conformal chart for
the unit disk $\mathbb{D}$ on the complex plane. Calling $z=\varrho
e^{i\phi}$ the new metric will be
\begin{equation}\label{eq:new_metric}
ds^{2} = w(z)(d\varrho^{2}+\varrho^{2}d\phi^{2})\,.
\end{equation}
The conformal factor $w(z)$ can be found by equating the metric
\eqref{eq:new_metric} with the original one. At this point it is
worth treating the problem in a slightly more general form. Consider
the case of a generic surface of revolution of the form:
\begin{equation}\label{eq:surface_of_revolution}
\left\{
\begin{array}{l}
x = \xi(r)\cos\phi\\
y = \xi(r)\cos\phi\\
z = \eta(r)
\end{array}
\right.,
\end{equation}
with $r \in [0,\infty[$ and $\phi \in [0,2\pi]$. The metric of the
surface \eqref{eq:surface_of_revolution} will be:
\begin{equation}\label{eq:old_metric}
ds^{2} = Edr^{2}+2Fdrd\phi+Gd\phi^{2},
\end{equation}
where
\begin{equation}
\left\{
\begin{array}{l}
E = \xi'^{2}+\eta'^{2}\\
F = 0\\
G = \xi^{2}
\end{array}
\right.
\end{equation}
are the coefficients of the first fundamental form of the surface
\eqref{eq:surface_of_revolution}. Equating \eqref{eq:old_metric} and
\eqref{eq:new_metric} one finds immediately:
\begin{equation}
w(\varrho) = \left[\frac{\xi(r)}{\varrho}\right]^{2},
\end{equation}
with $\varrho$ and $r$ related by the differential equation:
\begin{equation}
\frac{d\varrho}{dr}\pm\sqrt{\frac{E}{G}}\,\varrho = 0,
\end{equation}
whose solution is given by:
\begin{equation}\label{eq:conformal_map}
\varrho = \exp\left(\mp\int dr\,\sqrt{E/G}\right).
\end{equation}
The sign of the exponent and the integration constant in
\eqref{eq:conformal_map}, can be tuned to obtain the desired scale
and direction of the conformal map. It is easy to show that, in the
new coordinates, the Laplace operator takes the form:
\begin{equation}\label{eq:laplacian}
\Delta = w^{-1}\Delta_{z},
\end{equation}
in which $\Delta_{z}$ is the Laplacian in the Euclidean metric:
\begin{equation}\label{eq:euclidean_metric}
\gamma =
\left(
\begin{array}{cc}
1 & 0\\
0 & \varrho^{2}
\end{array}
\right).
\end{equation}
With this new set up, the Green equation
\eqref{eq:biharmonic_green_equation} can be easily written in the
form:
\begin{equation}\label{eq:conformal_biharmonic}
\Delta_{z} w^{-1} \Delta_{z}G_{2L}(z,\zeta) = \delta(z,\zeta),
\end{equation}
where $\zeta=r'e^{i\phi'}$ is a second generic point of the complex
plane and $\delta(z,\zeta)$ has the same meaning as in Sec.
\ref{sec:elastic_free_energy} with respect the Euclidean metric
$\gamma$. The differential operator $\Delta w^{-1}\Delta$ is known
in analysis as \emph{weighted} biharmonic operator. Analogously, the
Green-Dirichlet problem \eqref{eq:green_equation} becomes:
\begin{equation}\label{eq:conformal_green}
\left\{
\begin{array}{ll}
\Delta G_{L}(z,\cdot) = \delta(z,\cdot)    &\quad z\in\mathbb{D} \\[7pt]
G_{L}(z,\cdot)        = 0                      &\quad z\in\partial\mathbb{D}
\end{array}
\right.,
\end{equation}
so that:
\begin{equation}\label{eq:complex_green_function}
G_{L}(z,\zeta) = \frac{1}{2\pi}\log\left|\frac{z-\zeta}{1-z\overline{\zeta}}\right|.
\end{equation}
It must be stressed that so far we didn't explicitly use the
geometry of the paraboloid. What has been said, therefore, holds for
any surface of revolution which can be conformally mapped onto the
unit disk. Furthermore, as anticipated in Sec.
\ref{sec:elastic_free_energy}, the conformal distance $\varrho$,
which completely embodies the geometry of the surface, depends only
on the coefficients $E$ and $G$ of the first fundamental form.

In the particular case of the paraboloid we have:
\begin{subequations}
\begin{equation}
\varrho = \lambda\frac{re^{\sqrt{1+\kappa^{2}r^{2}}}}{1+\sqrt{1+\kappa^{2}r^{2}}},
\end{equation}
\begin{equation}
\lambda = \frac{1+\sqrt{1+\kappa^{2}R^{2}}}{R\exp\left(\sqrt{1+\kappa^{2}R^{2}}\right)}\cdot
\end{equation}
\end{subequations}
To obtain the expression for $\Gamma(\bm{x})$ given in
Eq.\eqref{eq:gamma_2} we are left with the task of calculating the
integral:
\begin{equation}\label{eq:screening_integral}
\Gamma_{s}(\bm{r}) = \Gamma_{s,1}(\bm{x})-\Gamma_{s,2}(\bm{r}),
\end{equation}
where
\begin{subequations}
\begin{align}
\Gamma_{s,1}(\bm{x}) &= \frac{1}{2\pi}\int d\phi'\,dr'\,\sqrt{g}\,K(r')\log|z-\zeta|             \label{eq:gamma_s1},\\
\Gamma_{s,2}(\bm{x}) &= \frac{1}{2\pi}\int d\phi'\,dr'\,\sqrt{g}\,K(r')\log|1-z\overline{\zeta}| \label{eq:gamma_s2}.
\end{align}
\end{subequations}
For this purpose we can use the expansion:
\begin{equation}\label{eq:log_expansion}
\log|z-\zeta| = \log\varrho_{>}-\sum_{n=1}^{\infty}\frac{1}{n}\left(\frac{\varrho_{<}}{\varrho_{>}}\right)^{n}\cos n\delta\phi,
\end{equation}
where $\varrho_{>}$ ($\varrho_{<}$) represents the largest
(smallest) modulus between $z$ and $\zeta$, while
$\delta\phi=\phi-\phi'$. The factorization of the angular variables
in Eq.\eqref{eq:log_expansion}, together with the pure radial
dependence of the Gaussian curvature and $\sqrt{g}$, makes the
angular dependence of $\Gamma_{s,1}$ vanish, so that we have:
\begin{align}
\Gamma_{s,1}(r)
&= \log\varrho(r)\int_{0}^{r} dr'\,\frac{\kappa^{2}r'}{(1+\kappa^{2}r'^{2})^{\frac{3}{2}}} \notag\\
&+ \int_{r}^{R} dr'\,\frac{\kappa^{2}r'}{(1+\kappa^{2}r'^{2})^{\frac{3}{2}}}\log\varrho(r'),
\end{align}
which integrated by parts gives:
\begin{equation}\label{eq:final_gamma}
\Gamma_{s,1}(r) = \log\left(\frac{\alpha e^{\sqrt{1+\kappa^{2}r^{2}}}}{1+\sqrt{1+\kappa^{2}r^{2}}}\right).
\end{equation}
Using an expansion similar to \eqref{eq:log_expansion} it is also
possible to prove that
\begin{equation}
\log|1-z\overline{\zeta}| = -\sum_{n=1}^{\infty}\frac{1}{n}\,(\varrho\varrho')^{n}\cos\delta\phi,
\end{equation}
which integrated over the surface of the paraboloid gives
$\Gamma_{s,2} = 0$. This last conclusion, combined with Eq.
\eqref{eq:final_gamma}, yields Eq.\eqref{eq:gamma_screening}.

\section{\label{sec:harmonic_kernel}The Harmonic Kernel}
In this appendix we give the exact expression for the harmonic
kernel appearing in Eq.\eqref{eq:harmonic_kernel}. This expression
has been found by Shimorin in the more general case of the
calculation of the Green function for the weighted biharmonic
operator with radial weight $w(z)=w(|z|^{2})$ \cite{Shimorin:1998}.
As in the previous appendix we call $z=\rho(r)e^{i\phi}$ and
$\zeta=\rho'(r')e^{i\phi'}$ two points of the complex plane that are
images of the points $(r,\phi)$ and $(r',\phi')$ of $\mathbb{P}$
under the conformal transformation \eqref{eq:rho}. The harmonic
kernel $H(z,\zeta)$ can be written in integral form as:
\begin{equation}
H(z,\zeta)
= -\int_{|\zeta|}^{1} \frac{dt}{\pi t} \int_{0}^{t} ds\,\sqrt{g}\,k\left(\frac{\varrho^{2}(s)}{t^{2}}\zeta\,\overline{z}\right),
\end{equation}
in which:
\begin{equation}
k(z\overline{\zeta})
= \sum_{n \ge 0}\frac{(z\overline{\zeta})^{n}}{c_{n}}+\sum_{n<0}\frac{(\overline{z}\zeta)^{|n|}}{c_{|n|}},
\end{equation}
where the coefficients $c_{n}$ are given by:
\begin{equation}
c_{n} = 2\int_{0}^{1}ds\,\sqrt{g}\,\varrho^{2n}(s).
\end{equation}

\section{\label{sec:differential_evolution}Optimization Via Parallel Differential Evolution}

\begin{figure}[t]
\centering
\includegraphics[scale=0.3,angle=-90,origin=c]{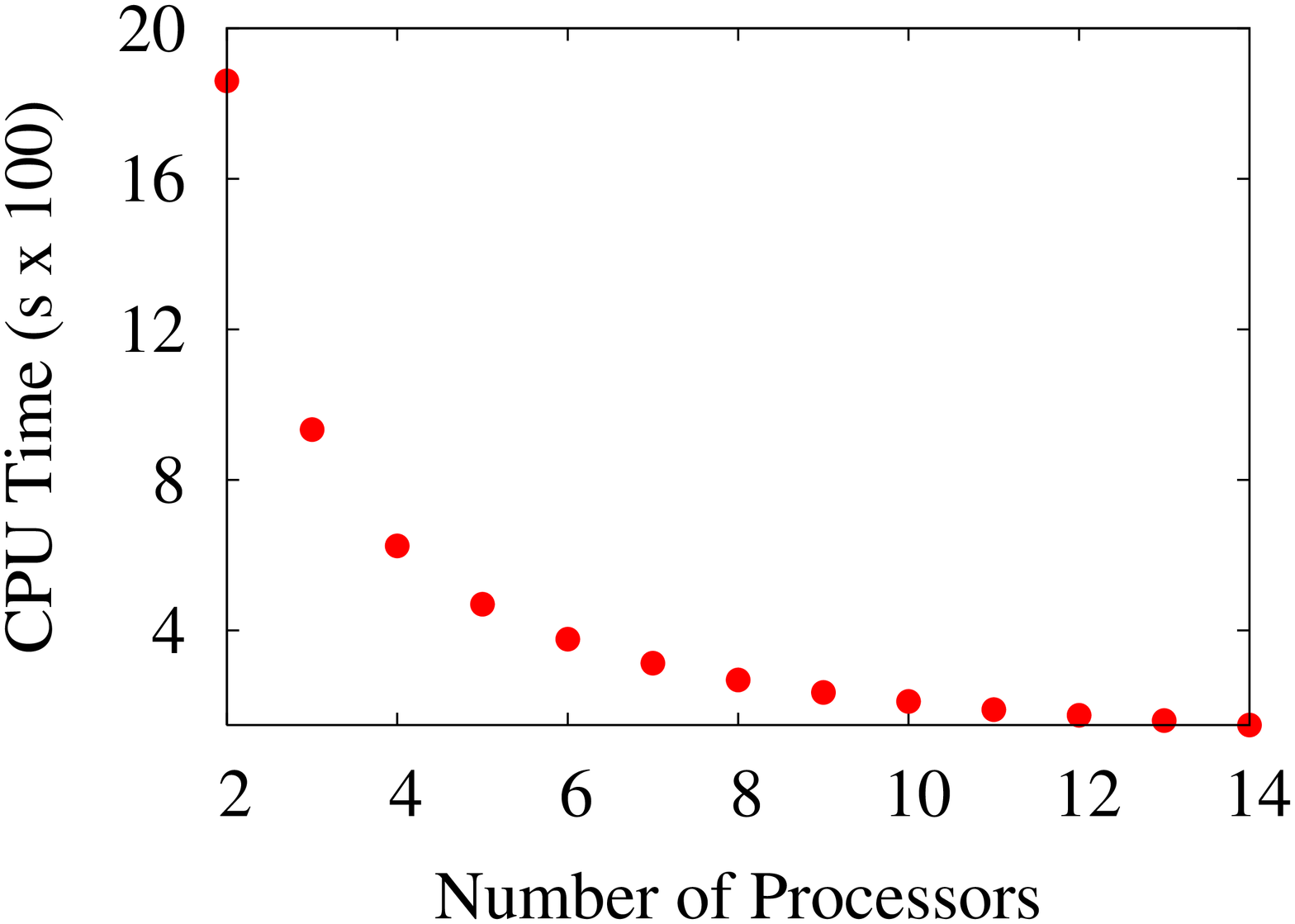}
\includegraphics[scale=0.3,angle=-90,origin=c]{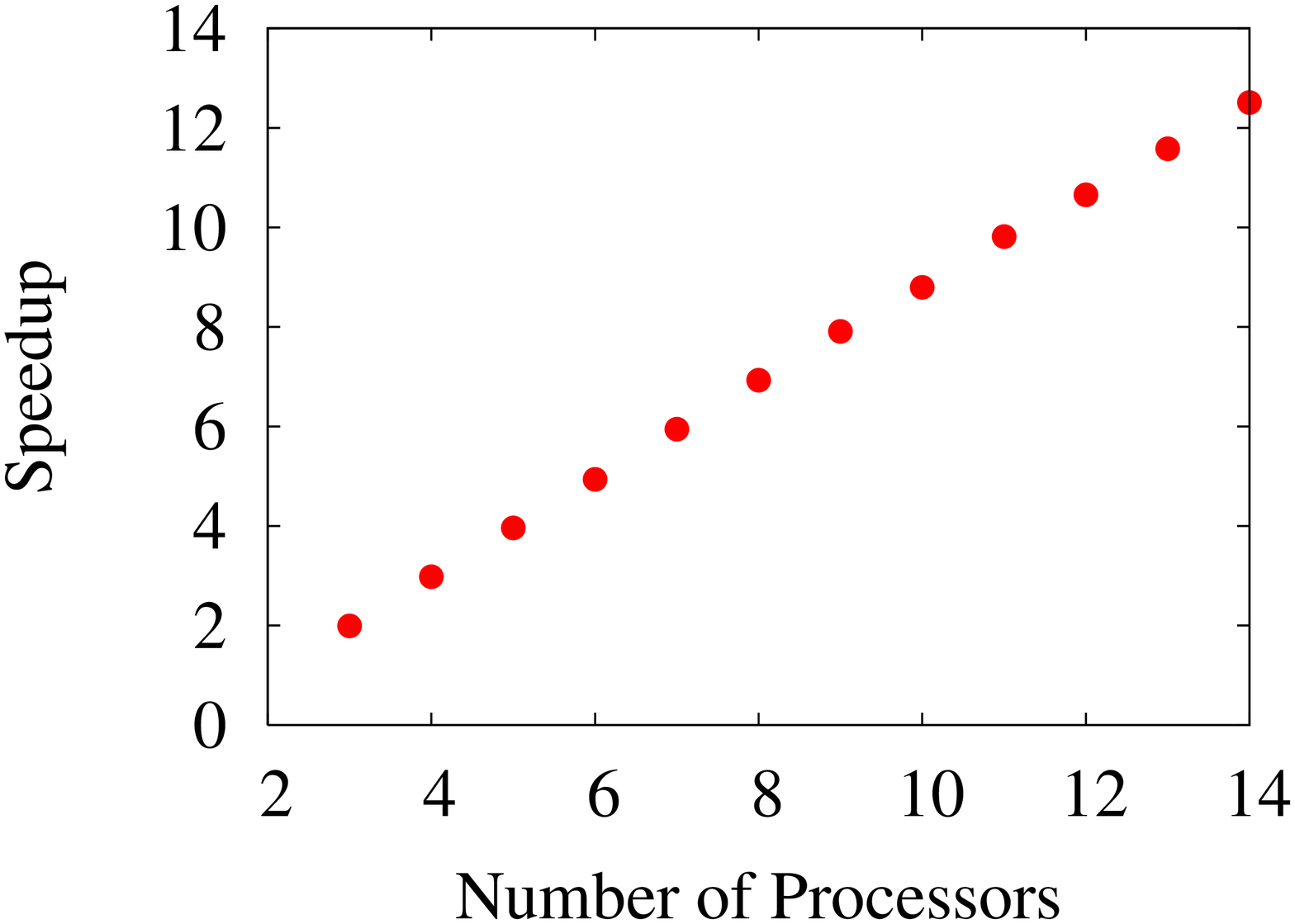}
\caption{\label{fig:parallel}CPU-Time and speedup (i.e. the time employed by $n$ processors to
accomplished a given number of iterations divided by the time employed by a
single machine to achieve the same task).}
\end{figure}

Many of the techniques proposed to determine the crystalline
structure of systems of interacting particles, as in the Thomson
problem, are based on local optimization procedures such as steepest
descent, conjugate gradient and the quasi-Newton method. Such
methods belong to the class of line-search algorithms for
multidimensional non-linear programming problems. They can be
described, in general, as a sequence of line minimizations of an
objective function along a set of directions that are generated
differently in different algorithms. Besides the well known local
convergence properties of these methods, they are generally unable
to locate the global minimum since they inherently approach the
closest local minimum for a given set of initial conditions.

To avoid the misconvergence problem described we adopt the
Differential Evolution (DE) algorithm of Storn and Price
\cite{StornPrice:1997}. This algorithm, which has been successfully
applied to several optimization problems in engineering
\cite{OnwuboluBabu}, belongs to the family of evolutionary
algorithms which are considerably faster than other stochastic
optimization methods, such as simulated annealing and genetic
algorithms, and more likely to find the correct global minimum.
These methods heuristically mimic biological evolution by
implementing natural selection and the principle of ``survival of
the fittest". An adaptive search procedure based on a population of
candidate solutions is used. Iterations involve a competitive
selection that drops the poorer solutions. The remaining pool of
candidates are perturbed (or mutated) in order to generate trial
individuals and then recombined with other solutions by a swap of
the components. The recombination and mutation moves are applied
sequentially; their aim is to generate new solutions that are biased
towards subsets of the search space in which good, although not
necessarily globally optimized, solutions have already been found.

An essential feature of Differential Evolution is the establishment
of genetic diversity, which helps to maximize the probability of
finding the true global minimum and to avoid misconvergence. One
begins with a large population of individuals uniformly distributed
in the search space. A good choice,  in practice, is to choose the
number of individuals to be an order of magnitude more than the
number of variables in the problem. The price one pays  is a
dramatic slowing down of the algorithm when applied to large scale
optimization. Considerable effort has therefore been made in the
past ten years to develop parallel implementations of evolutionary
algorithms aimed at reducing the overall time to completion of the
task by distributing the work on different processors working in
parallel. More recently some researchers have conjectured that some
parallelizations of a task improve the quality of the solution
obtained for a given overall amount of work (e.g. emergent
computation).

The \emph{Island Model} is a popular choice among parallelization
strategies and is implemented within a message passing model. It
consists of dividing the initial population into several
sub-populations and letting each of them evolve independently on a
single machine for a predetermined number of iterations (called the
\emph{epoch}). The exchange of genetic information is promoted by
swapping individuals between different sub-populations at the end of
each epoch. In the present work the migration strategy consists in
swapping the best individual of each sub-population with a randomly
selected individual on another island with the ring topology chosen
for the connectivity between islands. This choice allowed us to
achieve a substantial reduction of the CPU time and a linear speedup
(see Fig. \ref{fig:parallel}).

\bibliography{biblio}
\end{document}